\documentclass[11pt,a4paper]{article}
\usepackage{jheppub}

\usepackage{amsmath}
\usepackage{amsfonts}
\usepackage{amsthm}
\usepackage{slashed}
\usepackage{graphicx}
\usepackage{mathrsfs,bbm}
\usepackage{subfig}
\usepackage{booktabs}
\usepackage[numbers,sort&compress]{natbib}
\usepackage{lscape}
\usepackage{breqn}
\usepackage{amsmath,amssymb,textcomp,enumerate,alltt,xspace,multirow}
\usepackage{multicol}
\usepackage{boxedminipage}
\usepackage{comment}
\usepackage{tabularx}
\usepackage{adjustbox}

\usepackage[colorlinks=true]{hyperref}
\usepackage[dvipsnames]{xcolor}
\hypersetup{
colorlinks=true,
linkcolor=blue,
linktocpage=true,
citecolor=violet
}

\usepackage{multirow}
\allowdisplaybreaks

\makeatletter
\g@addto@macro\bfseries{\boldmath}
\makeatother

\theoremstyle{remark}

\DeclareMathOperator{\Li}{Li}
\DeclareMathOperator{\LS}{LS}

\DeclareMathOperator{\td}{d}
\newcommand{\dlog}{\mathrm{dlog}}
\newcommand{\iu}{\mathrm{i}}

\newcommand{\FI}{\mathcal{I}}
\newcommand{\TOP}[1][top]{\mathsf{#1}}
\newcommand{\PL}{\TOP[PL]}
\newcommand{\PLx}{\TOP[PLx12]}
\newcommand{\NP}{\TOP[NP]}
\newcommand{\FIPL}[1]{\FI_{\PL}(#1)}
\newcommand{\FINP}[1]{\FI_{\NP}(#1)}

\newcommand{\VVH}[1][V]{$#1 #1 H$}
\newcommand{\VVHH}[1][V]{$#1 #1 H H$}
\newcommand{\VVV}[1][V]{${#1}{#1}{#1}$}
\newcommand{\sbar}{s}
\newcommand{\tbar}{t}
\newcommand{\ubar}{u}
\newcommand{\mbar}{m}

\newcommand{\taumod}{y}

\newcommand{\liverpool}{Department of Mathematical Sciences, University of Liverpool, Liverpool L69 3BX, 
U.K.
}

\title{Two-loop light-quark Electroweak corrections to Higgs boson pair production in gluon fusion}

\author[a,b]{Marco Bonetti,}
\author[b]{Philipp Rendler,}
\author[c]{and William J.\ Torres~Bobadilla}

\affiliation[a]{Institute for Astroparticle Physics, Karlsruhe Institute of Technology, D-76344 Eggenstein \\Leopoldshafen, Germany}
\affiliation[b]{Institute for Theoretical Physics, Karlsruhe Institute of Technology, D-76128 Karlsruhe, Germany}
\affiliation[c]{\liverpool}

\emailAdd{marco.bonetti@kit.edu}
\emailAdd{philipp.rendler@kit.edu}
\emailAdd{torres@liverpool.ac.uk}

\abstract{
We compute two-loop electroweak corrections to double Higgs boson production in gluon fusion mediated by light quarks in a fully analytical way. We determine a basis of master integrals satisfying canonical differential equations in $\dlog{}$ form, enhanced by subsequent rotations to remove unnecessary functions that do not appear in the analytic expressions of the amplitudes. 
We determine the integration constants by matching our expressions to the large mass expansion limit of the canonical integrals.
We express the solution of differential equations in terms of Chen iterated integrals up to transcendental weight six over logarithmic kernels with algebraic arguments, and further decompose them by employing a basis of uniform weight functions. 
By deriving differential equations for such basis, we provide numerical results as well as routines for optimised numerical evaluations.
}


\preprint{{\raggedleft
            KA-TP-06-2025\\
            P3H-25-018\\
}}

\begin{document}

\maketitle

    \section{Introduction}
\label{sec:intro}

One of the main goals of the present and future runs of the Large Hadron Collider (LHC) at CERN is to strengthen the constraints on the trilinear Higgs boson self-coupling: starting from the present constraints on $\kappa$ factors on the self-coupling $\lambda_{HHH}$ of $-1.2<\kappa_\lambda<7.2$ from Atlas \cite{ATLAS:2024ish} and $-1.4<\kappa_\lambda<6.4$ from CMS \cite{CMS-PAS-HIG-20-011}, the High Luminosity phase of the LHC is expected to deliver the more stringent constraints of $0.1<\kappa_\lambda<2.3$, potentially excluding $\kappa_\lambda=0$ at the electroweak (EW) scale \cite{DiMicco:2019ngk}.

An important process entering the study of the trilinear Higgs self-coupling is double Higgs boson production from gluon fusion, where the trilinear coupling enters already at Leading Order (LO). This process has been extensively studied: LO results are available~\cite{Eboli:1987dy,Glover:1987nx}, and NLO QCD corrections with full top-quark mass dependence have also been calculated~\cite{Borowka:2016ehy,Borowka:2016ypz,Davies:2019dfy,Baglio:2018lrj,Baglio:2020ini,Campbell:2024tqg} and matched to parton showers~\cite{Jones:2017giv,Heinrich:2017kxx,Heinrich:2019bkc}, including anomalous couplings as well~\cite{Heinrich:2020ckp,Buchalla:2018yce,Bagnaschi:2023rbx,Heinrich:2022idm,Heinrich:2023rsd}. Higher-order contributions have been addressed in the heavy-top limit~\cite{DeFlorian:2018eng,deFlorian:2016uhr} or using expansions~\cite{Grigo:2015dia,Hu:2025aeo}, and partial three-loop results have become recently available~\cite{Davies:2021kex,Davies:2023obx,Davies:2024znp}. 
The NLO QCD results have been used as a base to include top-quark mass dependence at NNLO~\cite{Grazzini:2018bsd}, and at $\textup{N}^3\textup{LO}$ \cite{Chen:2019lzz,Chen:2019fhs} and $\textup{N}^3\textup{LO}+\textup{N}^3\textup{LL}$ \cite{AH:2022elh}, both in the heavy-top limit. 

Thanks to the inclusion of $\textup{N}^3\textup{LO}$ results, the residual scale uncertainty is estimated to be of about $3\%$, making the assessment of sub-leading contributions, such as EW corrections, a necessary theory ingredient. EW effects, despite being less impactful than top-mass renormalisation uncertainties (recently estimated to be around $4\%$ \cite{Heinrich:2019bkc,AH:2022elh,Jaskiewicz:2024xkd}), can produce relevant modifications in the shape of kinematic distributions and introduce interplays between QCD and EW renormalisation, making the point for their thorough investigation.

Partial results for EW corrections to double Higgs production in gluon fusion have been calculated~\cite{Borowka:2018pxx,Muhlleitner:2022ijf,Li:2024iio,Zhang:2024rix}, together with expressions in the large top-mass expansions~\cite{Davies:2023npk} and high-energy limit \cite{Davies:2022ram,Davies:2025wke}, while only recently a complete numerical evaluation of NLO EW corrections has been performed in~\cite{Bi:2023bnq}, followed by an independent computation of the class of diagrams featuring Higgs self-coupling and top-Yukawa interactions~\cite{Heinrich:2024dnz}.

In this work, we address the novel calculation of planar and non-planar four-point Feynman integrals with two on-shell and two off-shell external legs, and one internal mass.
We perform the analytic calculation of these integrals by following the method of differential equations in canonical form~\cite{Henn:2013pwa}, choosing a basis of integrals with uniform transcendental degree.
This class of Feynman integrals is particularly interesting to investigate, both because of its expected mathematical structure and for its phenomenological implications. On the mathematical side, these integrals are expected to be expressible in terms of iterated integrals over algebraic logarithmic kernels, in analogy to what it was observed for light-quark contributions to Higgs plus jet production~\cite{Bonetti:2020hqh,Bonetti:2022lrk}.
On the phenomenological side, light-quark contributions are a self-contained, gauge-invariant, and finite subset of the full NLO EW corrections to double Higgs production. In analogy with single Higgs results~\cite{Actis:2008ug,Degrassi:2004mx,Aglietti:2004nj}, we expect these contributions to be dominant when the Higgs couples to EW bosons only, thanks to the multiplicity of the light quarks and the lack of enhancement for top quark contributions coming from the Yukawa coupling. We expect this to be true both for the triangle-type diagrams (being the same as for single Higgs production) and for box-type diagrams, which are peculiar to double-Higgs production. Furthermore, QCD corrections to these contributions (i.e.\ \emph{mixed QCD-EW corrections}) can possibly increase the total cross section by a factor of $\mathcal{O}(+60\%)$, as observed in the NLO QCD-Yukawa case \cite{Borowka:2016ehy,Borowka:2016ypz,Baglio:2018lrj,Davies:2019dfy,Baglio:2020ini}.
In preparation for investigating mixed QCD-EW corrections, we produce expressions for the light-quark contributions up to order $\epsilon^2$.

Building on~\cite{Chicherin:2021dyp,Gehrmann:2024tds}, where a systematic method to construct a minimal basis of functions tailored to scattering amplitudes was developed, we organise the solution of master integrals according to their symbol alphabet to efficiently isolate functions with spurious features, ensuring they appear in the minimal number of basis elements. Once the mapping between the canonical integrals and the independent functions is found, we express the form factors in terms of these functions and construct differential equations without dependence on the dimensional regulator $\epsilon$, in a fully analytical fashion~\cite{Caron-Huot:2014lda,Badger:2021nhg}.
This organisation of integrals is beneficial for fast numerical evaluation and allows for analytically checking the cancellation of spurious singularities in the amplitudes.
In particular, we find remarkable simplifications in the calculation of finite contributions to the form factors presented in this paper, while higher orders in $\epsilon$ exhibit the expected complexity inherent to multi-loop computations.
The systematic treatment of independent canonical integrals and functions has been previously explored in the literature (see Refs.~\cite{Gehrmann:2018yef,Chicherin:2020oor,Chicherin:2021dyp,Badger:2021nhg,Badger:2023xtl,Abreu:2023rco,AH:2023ewe,Gehrmann:2024tds,Badger:2024gjs}).

Despite the complexity of the integration kernels, we integrate our differential equations up-to transcendental weight six in terms of Chen iterated integrals~\cite{Chen:1977oja}. All boundary constants are calculated in the large-mass expansion limit of the internal mass appearing in the integrals. 
For fast numerical evaluation of Feynman integrals and analytic expressions 
for the form factors, 
we make use of the method of generalised power series expansion, relying on the \textsc{Mathematica}
package {\sc DiffExp}~\cite{Moriello:2019yhu,Hidding:2020ytt}. With these numerical evaluation at hand, our analytic expressions are ready for phenomenological studies. \\

This paper is organised as follows.
In Sec.~\ref{sec:masters}, we determine a basis of canonical functions for the Feynman integrals, such that their differential equations are 
in canonical $\dlog$ form. 
We then explain how to express such canonical basis in terms of Chen iterated integrals up to transcendental weight six and construct a minimal basis of functions according to their symbol alphabet.
We determine boundary constants by matching the canonical functions to their large-mass expansion.
We devote Sec.~\ref{sec:amplitude} to describe the analytic structure of the amplitude and features of the form factors therein; 
furthermore, we construct differential equations for only the canonical integrals that are present in the analytic expressions of the form factors.
In Sec.~\ref{sec:results}, we present numerical results for the different parts of the amplitude. 
We draw our conclusions and future directions in Sec.~\ref{sec:conclusions}.
We accompany our paper with four appendices:
in Appendix~\ref{app:alphabet}, we list the logarithm kernels of the canonical differential equation; in Appendix~\ref{app:LME}, we present details on the analytic evaluation of boundary constants, obtained from the large-mass expansion limit; in Appendix~\ref{sec:appMATRIX}, we provide  details on the evaluation of the independent functions that appear in the form factor that contain the Feynman diagrams with $VVHH$ and $VVH$ vertices; in Appendix~\ref{app:cii}, we  discuss the numerical evaluation of Chen iterated integrals. 

~

The supplemental material of this paper, containing all the results of this work in \textsc{Mathematica} format, is provided at \cite{zenodo}.

\section{Canonical integrals for four-point functions with two off-shell legs and one internal mass}
\label{sec:masters}

\begin{figure}[t]
\centering
    \subfloat[$\FIPL{1,1,1,1,1,1,1,0,0}$]{{\quad\includegraphics[height=0.15\textheight]{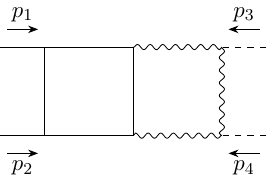}\quad}\label{fig:tsecLOPL}}
	\qquad\qquad
    \subfloat[$\FINP{1,1,1,1,1,1,1,0,0}$]{{\quad\includegraphics[height=0.15\textheight]{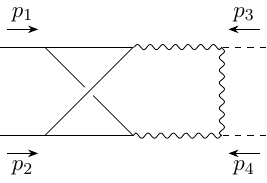}\quad}\label{fig:tsecLONP}}
	\caption{Two-loop planar ($\PL$) and non-planar ($\NP$) 
    integral families for four-point functions with two off-shell legs and one internal mass. Straight lines represent massless propagators or massless external legs, wavy lines propagators with mass $m_V$, and dashed lines external legs with $p^2 = m_H^2$.}
    \label{fig:tsecGG}
\end{figure}

In this paper we are interested in the analytic evaluation of four-point two-loop Feynman integrals
with two on-shell ($p_1^2=p_2^2=0$) and two off-shell ($p_3^2=p_4^2=m_H^2$) 
external legs with 
an internal mass, $m_V$. These integral families, depicted in Fig.~\ref{fig:tsecGG}, 
are described by the kinematic variables
\begin{subequations}
\begin{align}
 & s=(p_{1}+p_{2})^{2}\,, &  & t=(p_{1}+p_{3})^{2}\,, &  & u=(p_{2}+p_{3})^{2}\,,
\end{align}
that obey the momentum conservation relations 
\begin{align}
  p_{1}+p_{2}+p_{3}+p_{4}&=0\qquad\qquad\text{and}\qquad\qquad
  s+t+u=2m_H^{2}\,.  
\end{align}
\label{eq:kin_vars}
\end{subequations}
For the evaluation of these two-loop integrals, 
we adopt the normalisation
\begin{align}
\mathcal{I}^{(D)}_{\TOP[X]}(a_{1},a_{2},\dots,a_{9}) &= 
e^{2\epsilon\gamma_{\text{E}}}\,
\left({m_{V}^{2}}\right)^{2\epsilon}\,
\int\frac{\mathrm{d}^{D}k_{1}}{\iu\pi^{D/2}}\frac{\mathrm{d}^{D}k_{2}}{\iu\pi^{D/2}}\prod_{i=1}^{9}\frac{1}{D_{i}^{a_{i}}}\,,
\label{eq:Feyn_int}
\end{align}
with ${\TOP[X]}=\PL,\NP$, $D=4-2\epsilon$,
$\gamma_{\text E}$ the Euler--Mascheroni constant, and
$D_i$'s being loop ($i\leq7$) and auxiliary ($i=8,9$) propagators (summarised in Table~\ref{tab:toposGG}) with integer powers $a_i$. 

\begin{table}[t]
	\centering
	\begin{tabular}{cl@{\qquad}l@{\qquad}l}
\toprule
Denominator	&Integral family $\PL$			&Integral family $\NP$  \\
\midrule
$D_1$		&$k_1^2$					&$k_1^2$					\\
$D_2$		&$(k_1+p_1)^2$				&$(k_1+p_1)^2$				\\
$D_3$		&$(k_1-p_2)^2$				&$(k_1-k_2+p_2)^2$           \\	
$D_4$		&$(k_1-k_2)^2$				&$(k_1-k_2)^2$		        \\
$D_5$		&$(k_2+p_1)^2-m_V^2$		&$(k_2+p_1)^2-m_V^2$        \\
$D_6$		&$(k_2-p_2)^2-m_V^2$		&$(k_2-p_2)^2-m_V^2$        \\
$D_7$		&$(k_2+p_1+p_3)^2-m_V^2$	&$(k_2+p_1+p_3)^2-m_V^2$    \\
$D_8$		&$k_2^2$					&$k_2^2$					\\
$D_9$		&$(k_1+p_1+p_3)^2$			&$(k_1+p_1+p_3)^2$		    \\
\bottomrule
	\end{tabular}
	\caption{Definition of planar ($\PL$) and non-planar ($\NP$) integral families for the four-point functions depicted in Fig.~\ref{fig:tsecGG}. 
    The loop momenta are denoted by $k_1$ and $k_2$, while $m_V$ indicates the mass of the vector boson. The Feynman prescription $+ \iu\delta$ is understood for each propagator and not written explicitly.}
	\label{tab:toposGG}
\end{table}

\subsection{System of canonical differential equations for all integrals} 

For the analytic calculation of Feynman integrals, we follow the method of differential equations~\cite{Kotikov:1991pm, Bern:1993kr, Gehrmann:1999as}.
We opt for finding bases of pure transcendental functions that admit the canonical form~\cite{Henn:2013pwa}\footnote{To ease the notation, we drop the subscript $\TOP[X]$.} 
\begin{align}
\td\,\vec{J} & =\epsilon\,\td
\tilde{A}
\,\vec{J}\,,
\label{eq:deq_can}
\end{align}
with
\begin{align}
    \td&=\td s\,\frac{\partial}{\partial s}+\td t\,\frac{\partial}{\partial t}+\td u\,\frac{\partial}{\partial u}+\td m_{V}^{2}\,\frac{\partial}{\partial m_{V}^{2}}\,.
\end{align}
Explicitly, Eq.~\eqref{eq:deq_can} can be expressed in terms of partial derivatives as
\begin{align}
 & \frac{\partial\vec{J}}{\partial x}=\epsilon\,A_{x}\,\vec{J}\,,\quad\text{for }x\in\left\{ s\,,t\,,u\,,m_{V}^{2}\right\} \,,
 \label{eq:pdeqs}
\end{align}
where $A_{x}=\partial\tilde{A}/\partial x$.

When performing reductions from integration-by-parts identities (IBPs)~\cite{Chetyrkin:1981qh,Laporta:2000dsw}, we observe that the integral families $\PL$ and $\NP$ have, respectively, 
45 and 43 MIs.
We construct our canonical bases by looking for integrals that admit a $\dlog$ representation
with the aid of {\sc DlogBasis}~\cite{Henn:2020lye}. We rewrite the massive external momenta in terms of massless ones in order to employ the spinor-helicity formalism, which is well-suited for this package and allows us to fully benefit from its routines.
Interestingly, we observe that {\sc DlogBasis} performs remarkably well in constructing $\dlog$ integrals for the integral family $\NP$. 
However, for the family $\PL$, we need to first address each six-propagator sub-sector individually to automatically obtain $\dlog$ master integrals containing up to six propagators. 
This pattern is expected since the construction of $\dlog$ integrals strongly depends on the parametrisation of loop momenta. 
Only at this point we can address the construction of top sector integrals with seven propagators, using a loop-by-loop approach~\cite{Flieger:2022xyq}
in combination with additional rotations of the matrices $A_i$ given by Magnus transformation~\cite{Argeri:2014qva}.

The elements in the canonical basis contain kinematic prefactors, which are associated to the leading singularities of the Feynman integrals appearing therein. Such singularities contain both rational and algebraic functions, and in our case the latter are expressed in terms of the square roots:
\begin{subequations}
\begin{align}r_{1} & =\sqrt{s\left(s-4m_H^{2}\right)}\,,\\
r_{2} & =\sqrt{s\left(st^{2}-4m_{V}^{2}\left(m_H^{4}-tu\right)\right)}\,,\\
r_{3} & =\sqrt{s\left(m_{V}^{4}(s-4m_H^{2})+st^{2}-2m_{V}^{2}t(t-u)\right)}\,,\\
r_{4} & =\sqrt{s\left(s-4m_{V}^{2}\right)}\,,\\
r_{5} & =\sqrt{m_H^{2}\left(m_H^{2}-4m_{V}^{2}\right)}\,,\\
r_{6} & =\left.r_{2}\right|_{t\leftrightarrow u}=\sqrt{s\left(su^{2}-4m_{V}^{2}\left(m_H^{4}-tu\right)\right)}\,,\\
r_{7} & =\sqrt{s\left(m_H^{4}(s-4m_{V}^{2})+4m_{V}^{2}m_H^{2}(t+u)-4m_{V}^{2}tu\right)}\,,\\
r_{8} & =\left.r_{3}\right|_{t\leftrightarrow u}=\sqrt{s\left(m_{V}^{4}(s-4m_H^{2})+su^{2}+2m_{V}^{2}u(t-u)\right)}\,.
\end{align}
\label{eq:sqrts}
\end{subequations}
In particular, while the canonical basis for
family $\PL$ contains only the square roots $r_1$, $r_2$, $r_3$, $r_4$, and $r_5$,
in the basis for family $\NP$ all square roots appear. 
Let us note that under the $t\leftrightarrow u$ crossing
$r_7$ remains unchanged. This is expected, since this
square root comes from a genuine non-planar diagram
as opposed to the other ones.
We can associate the square roots listed in Eq.~\eqref{eq:sqrts} to the leading singularity of specific four-point loop integrals:
\begin{align}
&&\LS\left(\vcenter{\hbox{\includegraphics{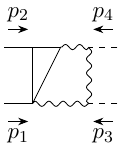}}}\right)
=r_2\,,
&& \LS\left(\vcenter{\hbox{\includegraphics{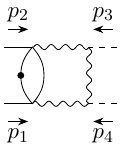}}}\right)
=r_3\,,
&&\LS\left(\vcenter{\hbox{\includegraphics{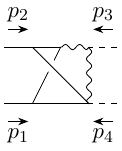}}}\right)
=r_7\,.
\label{eq:ls_sqrts}
\end{align}
Here the leading singularity $r_3$ can be understood from a loop-by-loop analysis,
where the one-loop bubble subtopology lies in $D=2-2\epsilon$ and the one-loop box in $D=4-2\epsilon$~\cite{Flieger:2022xyq}. 
Square roots $r_6$, $r_8$ are obtained by the exchange of $p_3\leftrightarrow p_4$, as mentioned above.
The remaining well known square roots $r_1$, and $r_4$  ($r_5$) come respectively from a 
three-point topologies with external momenta $\{p_1+p_2\,,p_3\,,p_4\}$,
and from one-loop bubbles with external momenta $\{p_1+p_2\,,p_3+p_4\}$ ($\{p_1+p_2+p_3\,,p_4\}$) and internal mass $m_V$. 

In order to obtain the canonical differential equation~\eqref{eq:deq_can}, we first construct the partial differential equations~\eqref{eq:pdeqs} using {\sc LiteRed}~\cite{Lee:2012cn} to compute derivatives and generate IBPs, and {\sc FiniteFlow}~\cite{Peraro:2019svx} to solve their linear system of equations and analytically reconstruct the matrices $A_x$ from numerical evaluations over finite fields~\cite{Klappert:2019emp,Peraro:2016wsq,vonManteuffel:2014ixa}. 

With the analytic expressions of all $A_x$ at our disposal, we proceed to determine $\tilde{A}$. 
We profit from the package {\sc Effortless}~\cite{Antonela}, which uses the even letters of the alphabet (derived from an analysis of Landau singularities)
and the square roots of Eq.~\eqref{eq:sqrts} to construct the algebraic letters.
These algebraic letters have the parametric form
\begin{align}
\frac{P-r_{i}}{P+r_{i}}\,, &  & \frac{Q-r_{i}\,r_{j}}{Q+r_{i}\,r_{j}}\,,
\end{align}
with $r_i$ square roots defined in Eq.~\eqref{eq:sqrts} and $P$, $Q$ holomorphic functions of the kinematic variables. 
By doing so, we obtain the almost complete alphabet defining these integral families.
The missing letters can easily be recovered by direct integration and have the form
\begin{align}
\frac{P-Q\,r_{i}}{P+Q\,r_{i}}\,.
\end{align}
We write an ansatz for the structure of $\tilde{A}$ utilising this alphabet, and fitting the matrix to match the values of each $A_x$.
Thus, the total differential of our canonical master integrals
takes the form
\begin{align}  
\td\vec{J}=\epsilon\,\sum_{i=0}^{77}\mathbb{A}_{i}\,\dlog\,\alpha_{i}\,\vec{J}\,,
\label{eq:deq_dlog}
\end{align}
where $\mathbb{A}_i$ are $\mathbb{Q}$-matrices (i.e.\ matrices of rational numbers), and $\alpha_i$'s are the letters of the alphabet, reported in Appendix~\ref{app:alphabet}.

\subsection{Explicit solution in terms of Chen iterated integrals}

We now proceed to integrate the canonical differential equations up to transcendental weight six.
We construct all integrals $\vec{J}$ as Laurent expansion in the dimensional regulator $\epsilon$ starting at $\mathcal{O}(\epsilon^0)$,
\begin{align}
\vec{J}\left(\vec{x};\epsilon\right)&=\sum_{k=0}^{6}\epsilon^{k}\,\vec{J}^{\left(k\right)}(\vec{x})+\mathcal{O}\left(\epsilon^{7}\right),
\label{eq:k_fold_ite}
\end{align}
where
$\vec{J}\left(\vec{x}\right)$ are transcendental functions of weight $k$, depending on the kinematic variables $\vec{x}=\{s,t,u,m_V^2\}$, given by the $k$-fold iterated integral
\begin{align}
\vec{J}^{\left(k\right)}\left(\vec{x}\right) & =\vec{J}^{\left(k\right)}\left(\vec{x}_{0}\right)+\int_{\gamma}\mathrm{d}\tilde{A}\,\vec{J}^{\left(k-1\right)}\left(\vec{x}\,'\right),
\label{eq:k_fold_chen}
\end{align}
with $\vec{J}^{\left(k\right)}\left(\vec{x}_{0}\right)$ corresponding to boundary
values at the base point $\vec{x}_0=\{s_0, t_0,u_0,m_{V;0}^2\} $, and 
$\gamma$ a path connecting the base point and another point $\vec{x}$. 

These integrals $\vec{J}^{\left(k\right)}$ can be written as
\begin{align}
\vec{J}^{\left(k\right)}\left(\vec{x}\right) & =\sum_{k'=0}^{k}\,\,\sum_{i_{1},\hdots.i_{k'}=0}^{77}
\mathbb{A}_{i_{1}}\hdots \mathbb{A}_{i_{k'}}\,\vec{J}^{\left(k-k'\right)}\left(\vec{x}_{0}\right)\left[\alpha_{i_{1}},\hdots,\alpha_{i_{k'}}\right]_{\vec{x}_{0}}\left(\vec{x}\right)\,,
\end{align}
where $\mathbb{A}_i$ corresponds to the $i$-th matrix associated to $\alpha_i$, 
according to the differential equation~\eqref{eq:deq_dlog}, 
and is recursively expressed
in terms terms of Chen iterated integrals~\cite{Chen:1977oja}
\begin{align}
\left[\alpha_{i_{1}},\hdots,\alpha_{i_{k}}\right]_{\vec{x}_{0}}(\vec{x})&=\int_{\gamma}\dlog\,\alpha_{i_{k}}(\vec{x}\,')\left[\alpha_{i_{1}},\dots,\alpha_{i_{k-1}}\right]_{\vec{x}_{0}}(\vec{x}\,')\,,
\end{align}
with $\left[\right]_{\vec{x}_{0}}=1$. 
Here the integration kernels depend on the letters of the alphabet $\vec{\alpha}$.
We refer the reader to Appendix~\ref{app:cii} for further details and numerical evaluation of Chen iterated integrals.

Before fully integrating out our differential equations
and committing with a particular phase-space region, 
we analyse the structure of the canonical basis. 
We carry out this study by looking at the symbol map
of the integrals $\vec{J}^{\left(k\right)}\left(\vec{x}\right)$~\cite{Goncharov:2010jf},
\begin{align}
\mathcal{S}\left[\vec{J}^{\left(k\right)}\left(\vec{x}\right)\right] & =\sum_{i_{1},\hdots.i_{k}=0}^{77}
\mathbb{A}_{i_{1}}\hdots \mathbb{A}_{i_{k}}\,\vec{J}^{\left(0\right)}\left(\vec{x}_{0}\right)\,\alpha_{i_{1}}\otimes\hdots\otimes\alpha_{i_{k}}\,,
\end{align}
which maps $k$-fold iterated integrals onto  $k$–fold tensor products. 
This operation allows us to understand at which weight a given letter starts appearing and then organise our canonical bases in terms of independent functions that manifest the dependence on particular integration kernels.
Notice that for the construction of the symbol only the boundary values of the weight zero function $\vec{J}^{(0)}(\vec{x}_0)$ are needed.\footnote{For more background material on this topic, see the recent review~\cite{Badger:2023eqz}.}
For instance, the integrals that appear in~\eqref{eq:ls_sqrts}
have the following symbol map, 
\begin{subequations}
\begin{align}
J_{\PL;17} & =-\Big(2\alpha_{10}\otimes\alpha_{37}+\alpha_{41}\otimes\alpha_{62}+\frac{1}{2}\alpha_{38}\otimes\alpha_{61}\Big)\epsilon^{2}+\mathcal{O}\left(\epsilon^{3}\right)\,,\\
J_{\NP;34} & =\Big(4\alpha_{8}\otimes\alpha_{44}-\alpha_{8}\otimes\alpha_{45}-\alpha_{8}\otimes\alpha_{46}-2\alpha_{8}\otimes\alpha_{47}-\alpha_{9}\otimes\alpha_{44}+\alpha_{9}\otimes\alpha_{45}+\alpha_{9}\otimes\alpha_{46}\nonumber \\
 & \quad+\alpha_{9}\otimes\alpha_{47}-2\alpha_{10}\otimes\alpha_{44}+\alpha_{10}\otimes\alpha_{45}+\alpha_{10}\otimes\alpha_{47}-2\alpha_{11}\otimes\alpha_{44}\nonumber \\
 & \quad+\alpha_{11}\otimes\alpha_{46}+\alpha_{11}\otimes\alpha_{47}\Big)\epsilon^2+\mathcal{O}\left(\epsilon^{3}\right)\,,\\
J_{\PL;36} & =\Big(\alpha_{8}\otimes\alpha_{12}\otimes\alpha_{34}+\frac{1}{2}\alpha_{8}\otimes\alpha_{1}\otimes\alpha_{35}+\frac{1}{2}\alpha_{8}\otimes\alpha_{33}\otimes\alpha_{50}-\frac{1}{4}\alpha_{38}\otimes\alpha_{61}\otimes\alpha_{57}\nonumber \\
 & \quad+\frac{3}{4}\alpha_{9}\otimes\alpha_{38}\otimes\alpha_{58}+\frac{3}{4}\alpha_{8}\otimes\alpha_{41}\otimes\alpha_{59}+\frac{3}{4}\alpha_{8}\otimes\alpha_{41}\otimes\alpha_{60}+\ldots\Big)\epsilon^{3}+\mathcal{O}\left(\epsilon^{4}\right)\,,
 \label{eq:J_PL_36}
\end{align}
\label{eq:symbol_Js}
\end{subequations}%
where the last entries of the symbols contain algebraic letters that depend on square roots;
ellipsis in Eq.~\eqref{eq:J_PL_36} contain similar symbols of weight three. 

\begin{table}
    \begin{center}
        \begin{small}
            \begin{tabular}{cc@{\qquad}c@{\qquad}c@{\qquad}c}
\toprule
Weight	            &$\PL$		            &$\PLx$			        &$\NP$                                 &Total\\
\midrule
1                   &2, 8, 9, 10, 38, 41    &2, 8, 9, 11, 38, 41    & 8, 9, 10, 11, 38, 41                  &7  \\
\midrule
\multirow{4}{*}{2}  &1, 3, 5, 6, 12,        &1, 4, 5, 6, 13,        & 1, 2, 3, 4, 5, 6, 12, 13,             &\multirow{4}{*}{33}  \\
                    &14, 18, 21, 23, 32,    &14, 18, 22, 24, 32,    &14, 21, 22, 23, 24, 25, 33, 37,        &   \\
                    &33, 37, 39, 52, 53,    &33, 40, 49, 52, 53,    &39, 40, 44, 45, 46, 47, 49, 52,        &   \\
                    &54, 61, 62, 74, 75     &54, 66, 70, 74, 75     &53, 54, 61, 62, 66, 70, 74, 75         &   \\
\midrule
\multirow{4}{*}{3}  &7, 17, 19, 28,         &7, 17, 19, 27,         &7, 17, 19, 20, 26, 27, 28, 29, 30,     &\multirow{4}{*}{35}  \\
                    &29, 31, 34, 35,        &30, 31, 42, 43,        &31, 34, 35, 36, 42, 43, 48, 50, 51,    &   \\
                    &36, 50, 51, 57,        &48, 55, 56, 64,        &55, 56, 57, 58, 59, 60, 63, 64, 65,    &   \\
                    &58, 59, 60, 73         &67, 68, 71, 72         &67, 68, 69, 71, 72, 73, 76, 77         &   \\
\midrule
4                   &13, 16, 63             &12, 15, 63             &15, 16                                 &2  \\
\bottomrule
            \end{tabular}
        \end{small}
    \end{center}
\caption{%
List of the letters appearing in the integral families $\PL$, $\PLx$ ($t\leftrightarrow u$ crossing of $\PL$), and $\NP$.
These letters are categorised according to the transcendental weight at which they first appear in the symbol.
All of the letters that appear at the symbol of transcendental weight $k-1$ also appear at weight $k$.
Notice that starting at weight 5 no new letters appear. 
}\label{tab:letters}
\end{table}

From Eqs.~\eqref{eq:symbol_Js}, we immediately appreciate at which transcendental weight a particular letter starts appearing. In details, 
letters $\alpha_{37}$, $\alpha_{44}$, $\alpha_{45}$, $\alpha_{46}$, $\alpha_{47}$, $\alpha_{61}$, and $\alpha_{62}$ 
start appearing at weight two, while letters 
$\alpha_{34}$, $\alpha_{35}$, $\alpha_{50}$, $\alpha_{57}$, $\alpha_{58}$, $\alpha_{59}$, and $\alpha_{60}$ 
appear at weight three. 
We present in Table~\ref{tab:letters} the transcendental weight at which the letters of the alphabet first appear in each integral family. This classification provides insight into the structure of the integrals and into the complexity of the functions involved in their evaluation. This organisation allows us for constructing differential equations in terms of independent functions, as we will describe in the following.

Several studies on the construction of independent functions
have been already performed in literature. In the following, we adopt the strategy of~\cite{Gehrmann:2024tds,Badger:2021nhg}.

\subsection{Independent functions}
\label{secsub:wfun}

Owing to the symbol map of the canonical integrals, we look for linear relations that these integrals satisfy at each transcendental weight. 
Order by order in transcendental weight, we construct a rotation such that the new elements of the canonical bases remain linearly independent. 
We have efficiently automated this procedure within the {\sc FiniteFlow} framework.
Explicitly, we construct the set of canonical integrals $\vec{W}$, 
\begin{align}
\vec{W}(\vec{x};\epsilon)&=R\,\vec{J}(\vec{x};\epsilon)\,,
\label{eq:rot_wfuns}    
\end{align}
with 
$R$ a $\mathbb{Q}$-matrix. 
In this way cancellations, expected to happen
once the $\epsilon$ expansion of master integrals 
is considered, are already accounted by $\vec{W}$. 
\begin{table}[t]
\centering
    \begin{tabular}{ccccc}
        \toprule
        {} & {$\PL$} & {$\PLx$} & {$\NP$} & {Total}\\
        \midrule
        {$\vec{W}_0$} & 1 (1) & 1 (0) & 1 (0) & 1\\
        {$\vec{W}_1$} & 6 (6) & 6 (1) & 6 (0) & 7\\
        {$\vec{W}_2$} & 17 (17) & 20 (6) & 19 (2) & 25\\
        {$\vec{W}_3$} & 15 (15) & 19 (9) & 16 (6) & 30\\
        {$\vec{W}_4$} & 6 (6) & 6 (4) & 1 (1) & 11\\
      \bottomrule
    \end{tabular}
\caption{
Relation of integral families $\PL\,,\PLx\,,$ and $\NP$ and canonical integrals $\vec{W}_k$. Numbers in parenthesis correspond to integrals that appear for the first time in this integral family. 
Notice that starting at weight 5 no new integrals appear. 
}\label{tab:W_can_ints}
\end{table}
We construct the set of 74 independent canonical integrals
by giving preference to planar over non-planar integrals.
In other words, and abusing the notation, we construct 
$\vec{J} = \vec{J}_\PL \cup \vec{J}_\PLx \cup \vec{J}_\NP$,
accounting from all symmetry relations between families.
We find with the rotation of Eq.~\eqref{eq:rot_wfuns} 
the decomposition of canonical integrals of families $\PL\,,\PLx\,,\NP$ in terms of integrals $\vec{W}_k$ 
with $k$ the transcendental weight at which this integral appears. 
In Table~\ref{tab:W_can_ints}, we provide a classification of 
the families $\PL$, $\PLx$, and $\NP$ in terms of the rotated integrals $\vec{W}_k$.

Because integrals $\vec{W}_k$ are constructed to be independent order-by-order in transcendental weight, we further decompose them in terms of independent functions,
\begin{align}
W_{i_{k}}\left(\vec{x};\epsilon\right) & =\sum_{k'=k}^{6}\epsilon^{k'}\,w_{i_{k}}^{\left(k'\right)}\left(\vec{x}\right)\,.
\label{eq:ind_wfun}
\end{align}
In this equation, the subscript $i_{k}$ accounts for the $i$-th
canonical integral that arises at transcendental weight $k$. As mentioned above,  $w_{i_{k}}^{\left(k'\right)} = 0$ for $k'<k$. 
The dimension of the complete set of functions $w_{i_{k}}^{\left(k'\right)}$ present in the integrals $\vec{W}$ is obtained as
\begin{align}
\dim\left(\left\langle\vec{w}^{\left(0\right)},\ldots,\vec{w}^{\left(k'\right)}\right\rangle\right)= & \sum_{k=0}^{\min\left(k',4\right)}\dim\left(\vec{W}_{k}\right)\,\left(1+k'-k\right)\,.
\label{eq:count_wfun}
\end{align}
In particular, up to transcendental weight four, five and six, we get 179, 253 and 327 independent functions, respectively. 
This decomposition motivates us to  
construct and solve differential equations for the
transcendental functions $w_{i_k}^{\left(k'\right)}$~\cite{Badger:2024gjs},
\begin{align}
\td w_{i_{k}}^{(k')} &= \sum_{l=0}^{6}
\td\Omega_{i_{k}\,i_{l}}\, w_{i_{l}}^{(k'-1)}\,,
\label{eq:dwfun}
\end{align}
which are independent of $\epsilon$, and whose kernels of integration, contained in the matrix $\Omega$, 
are the logarithmic forms encompassed by the alphabet $\vec{\alpha}$. 
This approach allows for lessening the number of operations when numerically evaluating Feynman integrals and results in a simpler expression for the scattering amplitude, thanks to intermediate cancellations.
In particular, 
some letters may appear in the explicit calculation of integrals but not in the final expression of the amplitude. 
Organising Feynman integrals in terms of these functions allows us to perform only the strictly necessary evaluations, having already removed the vanishing contributions.
Similarly, having direct access to the functional space of the amplitude level provides insights into the structure of its perturbative expansion.

We observe from Table~\ref{tab:letters} that at weight one our canonical basis depends only on the following transcendental functions:
\begin{align}
\label{eq:weight_1_funs}
 & w_{1_{1}}^{\text{(1)}}=2\log\left(\frac{m_{H}^{2}-2m_{V}^{2}-r_{5}}{m_{H}^{2}-2m_{V}^{2}+r_{5}}\right)\,,
 \\
 & w_{2_{1}}^{\text{(1)}}=4\log\left(m_{V}^{2}-t\right)\,, &  & w_{5_{1}}^{\text{(1)}}=2\log\left(\frac{s-2m_{V}^{2}-r_{4}}{s-2m_{V}^{2}+r_{4}}\right)\,,\nonumber \\
 & w_{3_{1}}^{\text{(1)}}=4\log\left(m_{V}^{2}-s\right)\,, &  & w_{6_{1}}^{\text{(1)}}=2\log(-s)\,,\nonumber \\
 & w_{4_{1}}^{\text{(1)}}=4\log\left(m_{V}^{2}-m_{H}^{2}\right)\,, &  & w_{7_{1}}^{\text{(1)}}=4\log\left(m_{V}^{2}-u\right)\,.\nonumber 
\end{align}
Both the analytic expressions and the definition of our canonical basis in terms of independent functions $w_{i_{k}}^{\left(k'\right)}$ are provided in the supplemental material.

\subsection{Boundary values}

Let us now turn our attention to the analytic calculation of boundary
values. We observe that the integrals appearing in $\PL$ and $\NP$ can systematically
be calculated in the limit $s,t,u\ll m_{V}^{2}$, often referred to
as the large-mass expansion. This limit serves to fix boundary constants for
the canonical integrals $\vec{J}^{\left(k\right)}\left(\vec{x}\right)$,
or alternatively for the independent transcendental functions $w_{i_{k}}^{\left(k'\right)}\left(\vec{x}\right)$. 
This boundary point can be understood as $\vec{x}_{0}=\{s_0,t_0,u_0,m_{V;0}^2\}=\{s,0,0,1\}$, with $\left|s\right|\ll 1$.

Let us concentrate on the integrals depicted in Fig.~\ref{fig:tsecGG}.
We observe that the only non-vanishing regions in the 
large-mass expansion limit are
$k_{1}^{2}\ll m_{V}^{2}\sim k_{2}^{2}$ 
and $k_{1}^{2}\sim k_{2}^{2}\sim\left(k_{1}-k_{2}\right)^{2}\sim m_{V}^{2}$. 
These regions, for the planar integral, can  be diagrammatically understood as 
\begin{subequations}
\begin{align}
    \left.\vcenter{\hbox{\includegraphics{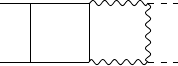}}}\right.
    &\;\to\;
    \left.\vcenter{\hbox{\includegraphics{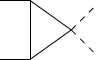}}}\right.
    \;\times\;
    \left.\vcenter{\hbox{\includegraphics{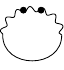}}}\right.
    \;+\; 
    \left.\vcenter{\hbox{\includegraphics{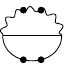}}}\right.\,,
\end{align}
whereas, for the non-planar integral,
one only has to account for the non-vanishing region 
$k_1^2\sim k_2^2\sim(k_1-k_2)^2\sim m_{V}^2$
\begin{align}
    \left.\vcenter{\hbox{\includegraphics{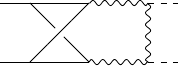}}}\right.
    \;\to\; 
    \left.\vcenter{\hbox{\includegraphics{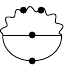}}}\right.\, .
\end{align}
\label{eq:diag_lme}
\end{subequations}%
We provide further details in Appendix~\ref{app:LME}. Here, we content ourselves with showing that, in the large-mass expansion, the only relevant integrals that require direct calculation are:
\begin{align}
 & \epsilon^{2}\,m_{V}^{2}\,\mathcal{I}_{\PL}^{\left(D-2\right)}\left(1,0,0,1,0,1,0,0,0\right)
 = \epsilon^2 m_V^2 \left(\vcenter{\hbox{\includegraphics{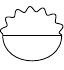}}}\right)^{(D-2)}\\
 & =-1-3\zeta_{2}\epsilon^{2}+\frac{8\zeta_{3}}{3}\epsilon^{3}-\frac{63\zeta_{4}}{4}\epsilon^{4}+8\left(\zeta_{2}\zeta_{3}+\frac{4\zeta_{5}}{5}\right)\epsilon^{5}-\left(\frac{32\zeta_{3}^{2}}{9}+\frac{869\zeta_{6}}{16}\right)\epsilon^{6}+\mathcal{O}\left(\epsilon^{7}\right)\,,\nonumber
\end{align}
and
\begin{align}
  \epsilon^{2}\,s\,\mathcal{I}_{\PL}^{\left(D-2\right)}
  &\left(0,1,1,0,0,0,1,0,0\right)
 = \epsilon^2 s \left(\left.\vcenter{\hbox{\includegraphics{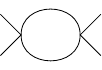}}}\right.\;\times\;\left.\vcenter{\hbox{\includegraphics{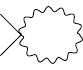}}}\right.\right)^{(D-2)}
 \\&
  =\left(\frac{m_V^2}{-s}\right)^{\epsilon}\left[-2+\frac{16\zeta_{3}}{3}\epsilon^{3}+6\zeta_{4}\epsilon^{4}+\frac{64\zeta_{5}}{5}\epsilon^{5}+\left(20\zeta_{6}-\frac{64\zeta_{3}^{2}}{9}\right)\epsilon^{6}\right]+\mathcal{O}\left(\epsilon^{7}\right)
  \nonumber\,.
\end{align}

Additionally, thanks to the rotation performed in~\eqref{eq:rot_wfuns}, all other integrals
vanish in this kinematic limit. 
This provides strong
evidence that such rotations significantly make the calculation of
integrals much more efficient, eliminating the need of providing boundary
constants for each canonical integral when integrating order-by-order
in $\epsilon$ (see Eqs.~\eqref{eq:k_fold_ite} and~\eqref{eq:k_fold_chen}).

In the boundary values, we observe a dependence on the kinematic variable
$s$. The presence of this variable does not introduce any ambiguity
in the calculation of boundary constants, as we can directly match
our expressions---formulated in terms of Chen iterated integrals---to
the boundary values once we account for the kinematic limit $s,t,u\ll m_{V}^{2}$.
The only subtlety to consider is the sign of $s$, since
$s=0$ represents a physical threshold, where one passes from
the unphysical ($s<0$) to physical ($s>0$) region. 

~

In the supplemental material of this paper, we provide analytic results
for the canonical bases in the physical regions $s>0$,
expressed in terms of Chen iterated integrals. 
It is important to
note that transitioning across other physical regions requires  appropriate
analytic continuation. In Sec.~\ref{sec:results}, 
in addition to the physical region
$s>0$, we draw our attention to the production region $s>4m_H^{2}$,
which is the relevant region for the scattering amplitude discussed
in following section.

\section{Two-loop scattering amplitudes for \texorpdfstring{$gg\to HH$}{gg -> HH}}
\label{sec:amplitude}

In this section, we construct the analytic expressions for the two-loop scattering amplitudes describing the light-quark EW corrections to double Higgs production in gluon fusion, using the set of Feynman integrals calculated in Sec.~\ref{sec:masters}.

\subsection{Scattering amplitudes and form factors}

We consider the scattering process
\begin{align}
    g(p_{1})+g(p_{2}) \to H(-p_{3})+H(-p_{4})\,,
\end{align}
with the same kinematic constraints as in Eq.~\eqref{eq:kin_vars}.
\begin{figure}
\centering
	\includegraphics[height=0.11\textheight]{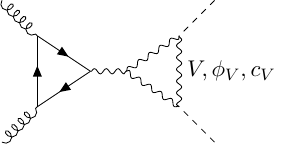}
	\caption{Example of two-loop light-quark factorisable diagram.}
    \label{fig:1l1l}
\end{figure}
\begin{figure}
\centering
    \subfloat[]{{\includegraphics[height=0.11\textheight]{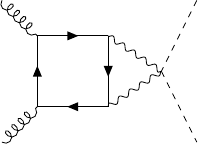}}\label{fig:HHVV}}
	\qquad\qquad
    \subfloat[]{{\includegraphics[height=0.11\textheight]{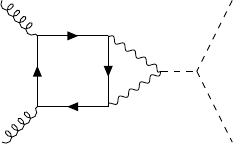}}\label{fig:VVHV}}
	\qquad\qquad
    \subfloat[]{{\includegraphics[height=0.11\textheight]{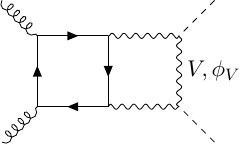}}\label{fig:VVV}}
	\caption{Representatives for the three different subsets of non-factorisable diagrams.}
    \label{fig:amplgg}
\end{figure}
All the quarks considered in this process are massless, therefore the Higgs boson can only couple to the EW bosons or to itself.\footnote{For text compactness, we refer to Goldstone bosons (if the gauge choice allows them) and to the $W$ and $Z$ bosons comprehensively as ``EW bosons''.} We always need a quark loop to connect the gluons to the EW part of the diagrams, which leads to the presence of a $\gamma^\mu \left(g_V + g_A \gamma_5\right)$ term from $Zq\overline{q}$ vertices. We can divide the diagrams contributing to the amplitude into two classes: \emph{factorisable diagrams}, which consist of two one-loop sub-diagrams connected via an EW boson line (e.g.\ Fig.~\ref{fig:1l1l}) and \emph{non-factorisable diagrams}, which feature genuine two-loop configurations (e.g.\ Fig.~\ref{fig:amplgg}). Since we are considering multi-loop integrals in dimensional regularisation, $\gamma_5$ is not properly defined and requires the application of a \emph{scheme} to be consistently handled. We adopt here the so-called Kreimer scheme~\cite{Kreimer:1989ke,Korner:1991sx}, see also discussion in~\cite[Section~2.1]{Chen:2024zju}.

The contribution of the factorisable diagrams is zero. At this level in perturbation theory, we can only have photon and $Z$ boson as connecting particles.\footnote{We cannot have Goldstone or Higgs bosons connecting the two sub-diagrams, since they couple to quarks proportionally to the quark masses.} Vector contributions vanish because of Furry theorem, while axial contributions vanish once we sum over complete generations of massless quarks \cite{Gehrmann:2015ora,vonManteuffel:2015msa}, as well as for charge-parity invariance of the sub-diagram connecting the vector boson and the two external Higgs particles.

The contribution of non-factorisable diagrams can be split into three different contributions, depending on how the Higgs boson pair is produced: diagrams containing a four-point vertex connecting two EW bosons and two Higgs bosons (dubbed \VVHH, cf.~Fig.~\ref{fig:HHVV}), diagrams where two EW bosons produce a single Higgs, which subsequently splits into two particles (\VVH, cf.~Fig.~\ref{fig:VVHV}), and diagrams where each Higgs boson is generated from a separate three-point vertex; as a consequence, these diagrams contain three massive lines with the same mass $m_V$ (\VVV, cf.~Fig.~\ref{fig:VVV}), when working either in unitary or in Feynman gauge. Furthermore, it is important to notice that in a single diagram either only $W$ bosons or $Z$ bosons (and the corresponding Goldstone particles) can be present. When considering a general $R_\xi$ gauge for the EW sector, the \VVHH{} and \VVH{} contributions contain 21 diagrams each, while the \VVV{} one contains 84 diagrams.

We stress that the light-quark contributions to $gg \to HH$ appear for the first time at two loops and represent a gauge-invariant, UV- and IR-finite set of diagrams. As a consequence, no renormalisation procedure is required.

In non-factorisable contributions, the vertices containing quarks and EW bosons are located on the same fermion loop. Thank to this, the amplitude consists of a part proportional to $(g_V^2 + g_A^2)$, not containing any $\gamma_5$, and of another part proportional to $g_V g_A$, containing a single $\gamma_5$. 

The part containing a single $\gamma_5$ vanishes due to charge-parity conservation once we sum over complete generations of massless quarks, as explained in~\cite{Gehrmann:2015ora,vonManteuffel:2015msa}. Since only the vector contribution remains, we can model the $Vq\overline{q}$ interaction with a vector coupling proportional to $\sqrt{g_V^2 + g_A^2}$.
The amplitude can be now decomposed into a linear combination of tensor structures identical to the one obtained for the Yukawa-QCD case.\footnote{No new tensor structures w.r.t.\ the QCD-Yukawa case are expected to appear when including internal axial couplings since we are considering the same set of external legs.} 

Thanks to Lorentz invariance, only two independent tensor structures are necessary to describe the amplitude~\cite{Borowka:2016ypz,Davies:2018ood,Davies:2018qvx,Davies:2019dfy},
\begin{align}
    \mathcal{M}_{\lambda_1 \lambda_2}^{c_1 c_2} = \delta^{c_1 c_2} \epsilon_{\lambda_1,\mu} \epsilon_{\lambda_2,\nu}
    \left[
        \mathcal{F}_1\, \mathbb{T}_1^{\mu\nu} 
        + \mathcal{F}_2\, \mathbb{T}_2^{\mu\nu}
    \right] ,
\end{align}
where $c_1$ and $c_2$ are the colour indices of the gluons and
\begin{align}
    \mathbb{T}_1^{\mu\nu} &= g^{\mu\nu} - \frac{1}{p_{12}}
        p_2^\mu p_1^\nu  \,,\\
    \mathbb{T}_2^{\mu\nu} &= g^{\mu\nu} + \frac{1}{p_{12} p_T^2}
    \left(
        m_H^2 p_2^\mu p_1^\nu - 2 p_{23} p_3^\mu p_1^\nu - 2 p_{13} p_2^\mu p_3^\nu + 2 p_{12} p_3^\mu p_3^\nu
    \right) ,
\end{align}
are tensor structures with,
\begin{align}
    p_T^2 = \frac{2 p_{13} p_{23}}{p_{12}} - m_H^2    \,,
\end{align}
and $p_{ij}=p_i\cdot p_j$. Notice also that $\mathbb{T}_{1,\mu\nu} \mathbb{T}_1^{\mu\nu} = \mathbb{T}_{2,\mu\nu} \mathbb{T}_2^{\mu\nu} = D-2$, and $\mathbb{T}_{1,\mu\nu} \mathbb{T}_2^{\mu\nu} = D-4$.

It is worth stressing that the above decomposition is independent of the gauge choice of the external gluons and allows for a direct identification of the form factors $\mathcal{F}_i$ with the helicity amplitudes of the process:
\begin{align}
    \begin{aligned}
        \mathcal{M}^{++} &= \mathcal{M}^{--} = -\mathcal{F}_1   \,,\\
        \mathcal{M}^{+-} &= \mathcal{M}^{-+} = -\mathcal{F}_2   \,.
    \end{aligned}
    \label{eq:ffha}
\end{align}
Since $W$ and $Z$ bosons never appear in the same diagram, the form factors can be decomposed as

\begin{align}
    \mathcal{F}_1 &= \mathcal{F}_{1,\textup{\VVV[W]}} + \mathcal{F}_{1,\textup{\VVHH[W]}} + \mathcal{F}_{1,\textup{\VVH[W]}} &&+ \mathcal{F}_{1,\textup{\VVV[Z]}} + \mathcal{F}_{1,\textup{\VVHH[Z]}} + \mathcal{F}_{1,\textup{\VVH[Z]}}
    \,,\\
    \mathcal{F}_2 &= \mathcal{F}_{2,\textup{\VVV[W]}} &&+ \mathcal{F}_{2,\textup{\VVV[Z]}}
    \,.
\end{align}
Each one of the $\mathcal{F}_{i,X}$ terms represents a gauge-invariant part of the amplitude. This property has been checked by explicitly computing the amplitude using either Feynman gauge or unitary gauge for the EW sector, and finding that the results are the same once the expressions have been written in terms of master integrals.

It is interesting to notice that the \VVH{} and \VVHH{} terms contribute only to $\mathcal{F}_1$, i.e.\ only to the $\mathcal{M}^{++}$ and $\mathcal{M}^{--}$ helicities; this is due to the fact that they present the same tensor structure as for single Higgs production, for which only $\mathcal{M}^{++}$ and $\mathcal{M}^{--}$ helicities are allowed.

The different parts of the form factors can be expressed in terms of just three scalar functions $\mathcal{A}_i\left(s/m_{V}^{2}\,,t/m_{V}^{2}\,,u/m_{V}^{2}\right)$ with $i=1,2,3$, of the kinematics:
\begin{subequations}
\begin{align}
\mathcal{F}_{1,\textup{\VVV}} & =\omega\,\tilde{G}_{V}\,\mathcal{A}_{1}\left(s/m_{V}^{2}\,,t/m_{V}^{2}\,,u/m_{V}^{2}\right)\,,\\
\mathcal{F}_{2,\textup{\VVV}} & =\omega\,\tilde{G}_{V}\,\mathcal{A}_{2}\left(s/m_{V}^{2}\,,t/m_{V}^{2}\,,u/m_{V}^{2}\right)\,,\\
\mathcal{F}_{1,\textup{\VVHH}} & =\omega\,\tilde{G}_{V}\,\mathcal{A}_{3}\left(s/m_{V}^{2}\right)\,,\\
\mathcal{F}_{1,\textup{\VVH}} & =\omega\,\tilde{G}_{V}\,\frac{3m_{H}^{2}}{s-m_{H}^{2}}\mathcal{A}_{3}\left(s/m_{V}^{2}\right)\,,
\end{align}
\label{eq:FFs_to_AAs}
\end{subequations}
for $V=W,Z$, where the functions $\mathcal{A}_i$ admit the QCD perturbative expansion
\begin{align}
    \mathcal{A}_i &= \sum_{L=2}^{\infty} a^{L-2}\,\mathcal{A}_i^{(L)}\,,
\end{align}
with
\begin{align}
    \begin{aligned}
    \omega & =-\iu(4\pi)^{2\epsilon}\textrm{e}^{-2\gamma_{E}\epsilon}\left(m_{V}^{2}\right)^{-2\epsilon}\frac{\alpha^{2}}{\sin^{4}\theta_{W}}\left(\frac{\alpha_S}{4\pi}\right),
\\
 a & =\iu(4\pi)^{\epsilon}\mathrm{e}^{-\gamma_{E}\epsilon}\left(\frac{m_{V}^{2}}{\mu^{2}}\right)^{-\epsilon}\left(\frac{\alpha_S}{4\pi}\right)
 \,,\\
    \tilde{G}_W   &= \sum_{\substack{i \in \{u,c\}\\j \in \{d,s,b\}}} \left|V_{ij}\right|^2
    \,,\\
    \tilde{G}_Z   &= \frac{1}{\cos^4 \theta_W} \sum_{q \in \{u,d,s,c,b\}} \left(g_{L,q}^2 + g_{R,q}^2\right)
                = \frac{1}{\cos^4\theta_W}\left(\frac{5}{4} - \frac{7}{3}\sin^2\theta_W + \frac{22}{9}\sin^4\theta_W\right),
    \end{aligned}
\end{align}
with $\mu$ being the regularisation parameter coming from higher order loop integrals, $\theta_W$ being the Weinberg angle and $V_{ij}$ being the Cabibbo--Kobayashi--Maskawa mixing matrix~\cite{Romao:2012pq,Denner:2019vbn}.\footnote{Assuming $V_{ij} = \delta_{ij}$ we retrieve the know result $\tilde{V}=2$, cfr.~\cite{Degrassi:2004mx,Bonetti:2016brm}.}

In general, to describe the unstable nature of the $W$ and $Z$ bosons, one needs to account for complex values for their masses.
This can be done by adopting  the complex mass scheme~\cite{Denner:2005fg}.
We nevertheless restrict ourselves to real values, since no resonant contributions appear in our process and we always consider phase-space points in the production region (which lies above any physical threshold). Moreover, our results are valid for arbitrary complex values of the variables, and can be evaluated by means of computer codes that allow for complex-valued inputs, such as \textsc{SeaSyde}~\cite{Armadillo:2022ugh} and \textsc{Line}~\cite{Prisco:2025wqs}.

The form factors $\mathcal{F}_1$ and $\mathcal{F}_2$ can be extracted by applying projectors:
\begin{align}
\mathcal{F}_i &=
    \mathbb{P}_{i,\mu\nu} 
    \left[
        \mathcal{F}_1\, \mathbb{T}_1^{\mu\nu} 
        + \mathcal{F}_2\, \mathbb{T}_2^{\mu\nu}
    \right]\,,
    \label{eq:ffs}
\end{align}
with
\begin{align}
    \begin{aligned}
        \mathbb{P}_{1,\mu\nu} &=
        \frac{1}{4}\frac{D-2}{D-3}\,\mathbb{T}_{1,\mu\nu} - \frac{1}{4}\frac{D-4}{D-3}\,\mathbb{T}_{2,\mu\nu}   \,,
        \\
        \mathbb{P}_{2,\mu\nu} &=
        \frac{1}{4}\frac{D-2}{D-3}\,\mathbb{T}_{2,\mu\nu} - \frac{1}{4}\frac{D-4}{D-3}\,\mathbb{T}_{1,\mu\nu}   \,.
    \end{aligned}
    \label{eq:proj}
\end{align}

\subsection{\texorpdfstring{$gg\to HH$}{gg -> HH} form factors at two loops}

\label{secsub:ffs}

We follow a standard procedure for the generation of the amplitude and the construction of the form factors. 
We produce Feynman diagrams relevant for the process with the computer code \textsc{Qgraf}~\cite{Nogueira:1991ex}, then extract the form factors applying the projectors $\mathbb{P}_1$ and $\mathbb{P}_2$ of Eqs.~\eqref{eq:proj}. We resolve the colour and Dirac algebra using the computer code \textsc{Form}~\cite{Ruijl:2017dtg}, writing the form factors as a linear combination of two-loop Feynman integrals (see Eq.~\eqref{eq:Feyn_int}).

We use the computer program~\textsc{Reduze}~\cite{vonManteuffel:2012np,fermatweb} to map these integrals (up to permutations of the external legs) onto the two integral families listed in Table~\ref{tab:toposGG} and depicted in~Fig.~\ref{fig:tsecGG}. We obtain a full symbolic reduction of the amplitude onto the basis of \emph{canonical integrals} 
discussed in Sec.~\ref{sec:masters} with the aid of the computer code \textsc{Kira}~\cite{Maierhoefer:2017hyi,Klappert:2019emp,fermatweb,Klappert:2020nbg,Klappert:2020aqs}.\footnote{%
Throughout the reduction process, we set $m_V^2=m_W^2=m_Z^2=1$. We reinstate their full dependence at the final stage by using dimensional analysis and keeping in mind that canonical integrals, as well as the amplitude in four dimensions, are dimensionless.}

In Sec.~\ref{sec:masters}, we mentioned that all integrals belonging to the families $\PL$, $\PLx$, 
and $\NP$ can be reduced to 74 canonical master integrals (see Table~\ref{tab:W_can_ints}). 
However, we notice that we can express $\mathcal{A}_1^{(2)}$ and $\mathcal{A}_2^{(2)}$ in terms of 72 canonical integrals, and $\mathcal{A}_3^{(2)}$ in terms of 9 integrals.
We summarise the number of independent functions and canonical integrals that appear in these form factors up to order  $\mathcal{O}\left(\epsilon^2\right)$ in Table~\ref{tab:F_VVHH}.

\begin{table}[t]
\centering
    \begin{tabular}{cccccccccc cccc c}
        \toprule
    \multirow{2}{*}{Weight} &&\multicolumn{3}{c}{$\mathcal{A}_1^{(2)}$}  &&\multicolumn{3}{c}{$\mathcal{A}_2^{(2)}$}  &&\multicolumn{3}{c}{$\mathcal{A}_3^{(2)}$} &&\multirow{2}{*}{Total}\\
        \cmidrule(lr){3-5}\cmidrule(lr){7-9}\cmidrule(lr){11-13}
        &&$\epsilon^0$  &$\epsilon^1$    &$\epsilon^2$ &&$\epsilon^0$ &$\epsilon^1$   &$\epsilon^2$ &&$\epsilon^0$ &$\epsilon^1$   &$\epsilon^2$ &  \\
        \midrule
		 $0$&&	0&	0&	0&&	0&	0&	0&& 1& 0& 0&& 1\\
		 $1$&&	4&	2&	0&&	6&	0&	0&& 2& 1& 0&& 7\\
		 $2$&&	19&	10&	2&&	25&	6&	0&& 1& 4& 1&& 32 \\
		 $3$&&	14&	35&	10&&	28&	27&	6&& 3& 1& 4&& 61\\
		 $4$&&	1&	22&	37&&	8&	31&	27&& --& 3& 1&& 66 \\
		 $5$&&	--&	1&	22&&	--&	8&	31&& --& --& 3&& 41\\
		 $6$&&	--&	--&	1&&	--&	--&	8&&  --& --& --&& 9\\
        \midrule
      {Canonical integrals} && 38 & 32 & 2 && 67 & 5 & 0 && 7& 2& 0&& 73\\
      \bottomrule
    \end{tabular}
\caption{
List of the functions and canonical integrals appearing in $\mathcal{A}_1^{(2)}$, $\mathcal{A}_2^{(2)}$, and $\mathcal{A}_3^{(2)}$.
Functions are categorised according to their transcendental weight at which they first appear 
at the given $\epsilon$ expansion of the form factors.
All of the functions of transcendental weight $k-1$ also appear at weight $k$.
}\label{tab:F_VVHH}
\end{table}

Similarly, in Table~\ref{tab:letters_wfun}, we present the letters of the alphabet $\vec{\alpha}$ that appear in the analytic expressions of the form factors, in terms of Chen iterated integrals. We classify them according to the transcendental weight of the functions $w_{i_k}^{(k')}$. 
This classification highlights the striking contrast in complexity when comparing the analytic evaluation of $\mathcal{A}_{3}^{(2)}$ against $\mathcal{A}_{1}^{(2)}$ and $\mathcal{A}_{2}^{(2)}$.

\begin{table}[t]
\centering
    \begin{footnotesize}
    \begin{tabular}{c c ccc c cccc c ccc}
        \toprule
    \multirow{2}{*}{Weight} &&\multicolumn{3}{c}{$\mathcal{A}_1^{(2)}$}  &&\multicolumn{3}{c}{$\mathcal{A}_2^{(2)}$}  &&\multicolumn{3}{c}{$\mathcal{A}_3^{(2)}$}\\
        \cmidrule(lr){3-5}\cmidrule(lr){7-9}\cmidrule(lr){11-13}
        &&$\epsilon^0$  &$\epsilon^1$    &$\epsilon^2$ &&$\epsilon^0$ &$\epsilon^1$   &$\epsilon^2$ &&$\epsilon^0$ &$\epsilon^1$   &$\epsilon^2$ &  \\
        \midrule
		 \multirow{2}{*}{1} &&	\multirow{2}{*}{8, 9, 10, 11}&	\multirow{2}{*}{2, 38}&	\multirow{2}{*}{--}&&	2, 8, 9, 10, &	\multirow{2}{*}{--}&	\multirow{2}{*}{--}&&	\multirow{2}{*}{9, 38}&	\multirow{2}{*}{--}&	\multirow{2}{*}{--}\\
         &&  &	&	&&	11, 38&	&	&&	&	&	\\
         \midrule
		 \multirow{8}{*}{2}&&	1, 2, 3, 4, 
&	\multirow{8}{*}{18, 32}&	\multirow{8}{*}{5}&&	1, 3, 4, 12, 
&	\multirow{8}{*}{5}&	\multirow{8}{*}{--}&&	\multirow{8}{*}{2}&	\multirow{8}{*}{5}&	\multirow{8}{*}{--}\\
        &&  12, 13, 14, 21, 
&	&	&&	13, 14, 18, 21, &	&	&&	&	&	\\

        &&  22, 23, 24, 25, 
&	&	&&	22, 23, 24, 25, &	&	&&	&	&	\\

        &&  33, 37, 38, 39, 
&	&	&&	32, 33, 37, 39, &	&	&&	&	&	\\

        &&  40, 41, 44, 45, 
&	&	&&	40, 41, 44, 45, &	&	&&	&	&	\\

        &&  46, 47, 49, 52, 
&	&	&&	46, 47, 49, 52, &	&	&&	&	&	\\

        &&  53, 54, 61, 62, 
&	&	&&	53, 54, 61, 62, &	&	&&	&	&	\\

        &&  66, 70, 74, 75  
&	&	&&	66, 70, 74, 75&	&	&&	&	&	\\
        \midrule
		 \multirow{9}{*}{3}&&	\multirow{2}{*}{5, 17, 19, 20,} 
&	\multirow{5}{*}{6, 7,} &	\multirow{9}{*}{--}&&	5, 7, 17, 19, 
&	\multirow{8}{*}{6, 26,} &	\multirow{9}{*}{--}&&	\multirow{9}{*}{5}&	\multirow{9}{*}{--}&	\multirow{9}{*}{--}\\

        &&	\multirow{2}{*}{27, 28, 31, 34,} &	\multirow{5}{*}{26, 29,} & &&	20, 27, 28, 29, & \multirow{8}{*}{69} &	&&	&	&	\\
        
        &&  \multirow{2}{*}{35, 36, 42, 43,} &	\multirow{5}{*}{30, 51,}   &   &&	30, 31, 34, 35, &&	&&	&	&	\\
        
        &&  \multirow{2}{*}{48, 50, 55, 57,} &	\multirow{5}{*}{56}   &   &&	36, 42, 43, 48, &&	&&	&	&	\\
        
        &&  \multirow{2}{*}{58, 59, 60, 63,}  &	&   &&	50, 51, 55, 56, &&	&&	&	&	\\
        
        &&  \multirow{2}{*}{64, 65, 67, 68,}   &	&   &&	57, 58, 59, 60,  &&	&&	&	&	\\
        
        &&  \multirow{2}{*}{69, 71, 72, 73,}    &	&   &&	63, 64, 65, 67,  &&	&&	&	&	\\
        
        &&  \multirow{2}{*}{76, 77}    &	&   &&	68, 71, 72, 73, &&	&&	&	&	\\
        
        &&      &	&   &&	76, 77  &&	&&	&	&	\\
        \midrule
		 4&&	--&	15, 16&	--&&	--&	15, 16&	--&&	--&	--&	--\\
        \midrule
      Total &&	66&	77&	77&&	72&	77&	77&&	4&	4&	4\\  
      \bottomrule
    \end{tabular}
    \end{footnotesize}
\caption{
List of the letters (integration kernels) appearing in the functions $\mathcal{A}_{1}^{(2)}\,,\mathcal{A}_{2}^{(2)}\,,\mathcal{A}_{3}^{(2)}\,,$ up-to $\mathcal{O}\left(\epsilon^2\right)$. 
Similar to Table~\ref{tab:letters}, we categorise letters according to the transcendental weight at
which they first appear in the iterated integral. 
All of the letters that appear at 
transcendental weight $k-1$ also appear at weight $k$. 
Starting at weigh five, no new letters appear. 
The last row contains that the total number of letters present in the indicated function. 
}\label{tab:letters_wfun}
\end{table}

In the following, we separately discuss each two-loop gauge invariant group.
Since the form factors involving internal $Z$ or $W$ bosons share the same analytic structure, as shown in Eqs.~\eqref{eq:FFs_to_AAs}, we concentrate on the two-loop contributions to $\mathcal{A}_1$, $\mathcal{A}_2$, and $\mathcal{A}_3$.

\subsubsection*{\texorpdfstring{\VVHH{}}{VVHH} and \texorpdfstring{\VVH{}}{VVH} contributions}

The analytic expressions for the function $\mathcal{A}_3^{(2)}$ (up to overall kinematic prefactors) has
already been obtained in the literature for single Higgs production
up to $\epsilon^{2}$, expressed in terms of of generalised polylogarithms~\cite{Goncharov:1998kja} up to transcendental weight five, due to a weight drop in the transcendental degree~\cite{Bonetti:2016brm}. 
With
our canonical integrals, we recompute these form factors  in terms of independent transcendental functions. 

We obtain the following expression in terms of nine canonical integrals $\vec{W}$:\footnote{Ref.~\cite{Bonetti:2016brm} reports the same number of integrals.
}
\begin{align}
\mathcal{A}_3^{(2)}= &
\left[-2\left(33\epsilon^{2}+7\epsilon+1\right)W_{1_{0}}-\frac{(s-1)}{4s}\left(\frac{1}{\epsilon}+10+55\epsilon\right)W_{3_{1}}\right.\nonumber \\
 & -\frac{(s-4)}{4r_{4}}\left(\frac{1}{\epsilon}+6+27\epsilon\right)W_{5_{1}}-\frac{1}{s}\left(-\frac{2}{\epsilon^{2}}+\frac{2s-11}{\epsilon}+15(s-2)\right)W_{5_{2}}\nonumber \\
 & -\frac{(s-4)}{4r_{4}}\left(\frac{2}{\epsilon}+11\right)W_{11_{2}}+\frac{(s-2)}{s}\frac{1}{\epsilon}W_{13_{2}}\nonumber \\
 & +\frac{1}{s}\left(\frac{s-2}{\epsilon^{3}}+\frac{4s-10}{\epsilon^{2}}+\frac{15s-36}{\epsilon}\right)W_{12_{3}}\nonumber \\
 & +\frac{(s-4)}{4r_{4}s}\left(\frac{s-2}{\epsilon^{3}}+\frac{4(s-2)}{\epsilon^{2}}+\frac{15s-32}{\epsilon}\right)W_{29_{3}}\nonumber \\
 & \left.-\frac{1}{2s}\left(\frac{s-4}{\epsilon^{3}}+\frac{4(s-3)}{\epsilon^{2}}+\frac{5(3s-8)}{\epsilon}\right)W_{30_{3}}\right]
 +\mathcal{O}\left(\epsilon^{3}\right),
 \label{eq:A3}
\end{align}
where we have included only the relevant coefficients of the canonical integrals $W_{i_{k}}$ (see Eq.~\eqref{eq:rot_wfuns}) and, for the sake of simplicity, we have set $m_V^2=1$. 
This structure of the amplitude follows from how the canonical integrals $W_{i_k}$
have been defined in terms of independent functions $w_{i_k}^{(k')}$ according to Eq.~\eqref{eq:ind_wfun}.

In particular, the finite part of $\mathcal{A}_3^{(2)}$ reads
\begin{align}
\left.\mathcal{A}_3^{(2)}\right|_{\epsilon^{0}}
&=2-\frac{(s-1)}{s}\log(1-s)+\frac{(s-4)}{2r_{4}}\log\left(\frac{2-s-r_{4}}{2-s+r_{4}}\right)-\frac{1}{s}\Li_2 s+\frac{(s-4)}{2s}\Li_3 s
\notag\\
&+\frac{(s-2)}{s}w_{12_3}^{(3)}+\frac{(s-4)(s-2)}{4r_{4}s}w_{29_3}^{(3)}\,,
\end{align}
where we have plugged in the decomposition of $W_{i_k}$, evaluated in the physical region $s>0$. The single $w_{i_k}^{(k')}$ appearing in the expression are
\begin{align}
  & w_{1_{0}}^{(0)}=-1\,, &  & w_{5_{2}}^{(2)}=-\frac{1}{2}\Li_{2}s\,, &  & w_{30_{3}}^{(3)}=-\Li_{3}s\,,
\end{align}
with weight one functions reported in Eq.~\eqref{eq:weight_1_funs}, and
\begin{align}
w_{12_3}^{(3)} & =\frac{3}{8}\left[\alpha_{9},\alpha_{38},\alpha_{38}\right]-\frac{1}{4}\left[\alpha_{38},\alpha_{38},\alpha_{2}\right]+\frac{1}{2}\Li_3 s\,,\nonumber \\
w_{29_3}^{(3)} & =2\left[\alpha_{9},\alpha_{2},\alpha_{38}\right]-3\left[\alpha_{9},\alpha_{38},\alpha_{5}\right]+\frac{1}{12}\log^{3}\frac{2-s-r_4}{2-s+r_4}\,.
    \label{eq:wkA3}
\end{align}
We observe that only seven independent functions (or canonical integrals) appear
in the finite part. This represents an improvement compared to the
initial number of master integrals required for the calculation. Explicitly,
the canonical integrals $W_{11_{2}}$ and $W_{13_{2}}$ first appear at $\mathcal{O}\left(\epsilon\right)$. 

Owing to the simplicity of $\mathcal{A}_3^{(2)}$, depending on only one variable,
the functions expressed in terms of Chen iterated integrals can be
readily converted into generalised polylogarithms by parametrising
$s$ along the path $s=\left(1+x\right)^{2}/x$ (with
$x\in]0,1]$).
This change of variable allows for the systematic numerical
evaluation of the form factors up to $\mathcal{O}\left(\epsilon^{2}\right)$
using available computer codes, such as \textsc{GiNaC}~\cite{Vollinga:2004sn}.

~

In addition to evaluating Chen iterated integrals or parametrising
them so that they can be expressed in terms of generalised polylogarithms,
we also take advantage of having analytic expressions for the
form factors in terms of independent functions to construct and solve differential
equations solely for these functions, as explained in Sec.~\ref{secsub:wfun}.
This task is straightforward, thanks to the \textsc{Mathematica} package \textsc{DiffExp}~\cite{Moriello:2019yhu,Hidding:2020ytt},
which requires only the differential equations satisfied by these
functions and their boundary values as input.

Let us consider the transcendental functions $w_{i_k}^{k'}$ appearing in Eq.~\eqref{eq:wkA3}. 
We can construct the differential equation,
\begin{align}
    \td\vec{\omega}_{3;0} = \td\Omega_{3;0}\,\vec{\omega}_{3;0}\,,
\end{align}
with the matrix of coefficients, according to Eq.~\eqref{eq:dwfun}, 
\begin{align}
\Omega_{3;0} &=
\begin{pmatrix}
 0 & 0 & 0 & 0 & 0 & 0 & 0 & 0 & 0 \\
 -4 L_9 & 0 & 0 & 0 & 0 & 0 & 0 & 0 & 0 \\
 2 L_{38} & 0 & 0 & 0 & 0 & 0 & 0 & 0 & 0 \\
 0 & \frac{L_2}{8} & 0 & 0 & 0 & 0 & 0 & 0 & 0 \\
 0 & -\frac{3}{4}L_{38} & 0 & 0 & 0 & 0 & 0 & 0 & 0 \\
 0 & 0 & \frac{L_{38}}{4} & 0 & 0 & 0 & 0 & 0 & 0 \\
 0 & 0 & 0 & -L_2 & -\frac{L_{38}}{8} & \frac{L_2}{2} & 0 & 0 & 0 \\
 0 & 0 & 0 & 4 L_{38} & L_5 & -L_{38} & 0 & 0 & 0 \\
 0 & 0 & 0 & 2 L_2 & 0 & 0 & 0 & 0 & 0 \\
\end{pmatrix}
\,,
\label{eq:mat_omega_30}
\end{align}
where $L_{i}\equiv\log\alpha_{i}$, and the 
basis $\vec{\omega}_{3;0}$ reads, 
\begin{align}
\vec{w}_{3:0}
=
\left\{ w_{1_{0}}^{\text{(0)}},w_{3_{1}}^{\text{(1)}},w_{5_{1}}^{\text{(1)}},w_{5_{2}}^{\text{(2)}},w_{11_{2}}^{\text{(2)}},w_{13_{2}}^{\text{(2)}},w_{12_{3}}^{\text{(3)}},w_{29_{3}}^{\text{(3)}},w_{30_{3}}^{\text{(3)}}\right\}\,.
\end{align}
with the only non-vanishing boundary value at $s=0$, 
\begin{align}
    \left.w_{1_{0}}^{\text{(0)}}\right|_{s=0} = -1\,.
\end{align}
Notice that the functions $w_{11_{2}}^{\text{(2)}},w_{13_{2}}^{\text{(2)}}$, 
which are required to construct the differential equation, do not
appear in the finite contribution to $\mathcal{A}_3^{(2)}$.
This is expected, since weight three functions are iterated integrals of weight two
ones.

In order to compute $\mathcal{A}_3^{(2)}$ up to $\mathcal{O}\left(\epsilon^{2}\right)$, we need to consider differential equations for 27 independent functions (see Appendix~\ref{sec:appMATRIX} for details), with the same kernels of integration as in $\Omega_{3;0}$. In Sec.~\ref{sec:results}, we numerically solve these differential equations across different physical regions.

\subsubsection*{\texorpdfstring{\VVV{}}{VVV} contributions}

The analytic calculation of the functions $\mathcal{A}_1^{(2)}$ and $\mathcal{A}_2^{(2)}$ represents the main result of this work,
derived from the novel computation of genuine four-point Feynman integrals
shown in Fig.~\ref{fig:tsecGG}. Due to the large alphabet characterising
these integral families, we do not attempt to rationalise the square roots
through a parametrisation of the kinematic variables, as performed
for $\mathcal{A}_3^{(2)}$. Instead, we express the solution of the differential equations in terms of Chen iterated integrals, in this way keeping full dependence on even and algebraic letters of the alphabet. 

The analytic expressions for $\mathcal{A}_1^{(2)}$ and $\mathcal{A}_2^{(2)}$ 
at $\mathcal{O}\left(\epsilon^0\right)$, $\mathcal{O}\left(\epsilon^1\right)$,
and $\mathcal{O}\left(\epsilon^2\right)$ are expressed
in terms of transcendental functions of up to weight four, five and six, respectively.
Differently from $\mathcal{A}_3^{(2)}$, no transcendental weight drop occurs.

Let us elaborate on the structure of the form factors in terms of independent functions $w_{i_k}^{(k')}$. Eq.~\eqref{eq:count_wfun} provides the number of functions appearing in the canonical integrals of families $\PL$, $\PLx$, and $\NP$ at each transcendental weight. Through the direct calculation of the form factors, we observe the presence of a smaller set of functions: the form factors can be expressed up to $\mathcal{O}\left(\epsilon^0\right)$, $\mathcal{O}\left(\epsilon^1\right)$, and $\mathcal{O}\left(\epsilon^2\right)$ in terms of 70, 142, and 214 functions, respectively. This represents a significant improvement compared to using the canonical integrals. We will take advantage of this approach in Sec.~\ref{sec:results}.

Tables~\ref{tab:F_VVHH} and~\ref{tab:letters_wfun} also show remarkable simplifications in the finite contributions to $\mathcal{A}_1^{(2)}$ and $\mathcal{A}_2^{(2)}$. $\mathcal{A}_1^{(2)}$, related to the helicity amplitude $\mathcal{M}^{(2)\,++}$, consists of 66 letters and only 38 independent functions, thanks to the inherent symmetries of the all-plus helicity configuration. $\mathcal{A}_2^{(2)}$, associated with $\mathcal{M}^{(2)\,+-}$, presents a more intricate structure, due to the lack of symmetry of the external states.

When considering the expressions of $\mathcal{A}_1^{(2)}$ and $\mathcal{A}_2^{(2)}$ in terms of uniform weight functions and up to $\mathcal{O}\left(\epsilon^{n}\right)$, we notice that only six out of the seven logarithms listed in Eq.~\eqref{eq:weight_1_funs} appear. We can then write
\begin{align}
\mathcal{A}_{i;n}^{\left(2\right)} & =c_{i;2}^{\left(n\right)}w_{2_{1}}^{\text{(1)}}+c_{i;3}^{\left(n\right)}w_{3_{1}}^{\text{(1)}}+c_{i;4}^{\left(n\right)}w_{4_{1}}^{\text{(1)}}+\frac{c_{i;5}^{\left(n\right)}}{r_{4}}w_{5_{1}}^{\text{(1)}}+c_{i;6}^{\left(n\right)}w_{6_{1}}^{\text{(1)}}+c_{i;7}^{\left(n\right)}w_{7_{1}}^{\text{(1)}}+\ldots
\end{align}
with $i=1,2$, $c_{i;j}^{(n)}$ rational functions depending on the kinematic variables and appearing at $\mathcal{O}\left(\epsilon^{n}\right)$, and ellipsis accounting for higher transcendental weight functions. Note that here no $w_{0_{1}}^{\text{(0)}}$ or $w_{1_{1}}^{\text{(1)}}$ are present, and we observe as well the absence of the canonical integrals $\vec{W}_0$ and $\vec{W}_1$ in the $D$-dimensional $\mathcal{A}_1^{(2)}$ and $\mathcal{A}_2^{(2)}$ functions written in the $\vec{W}$ basis of canonical integrals.\footnote{$w_{1_{0}}^{\text{(0)}}$ (and the canonical integral $\vec{W}_{1_0}$) appears in $\mathcal{A}_3^{(2)}$, as shown in Eq.~\eqref{eq:A3}, while $w_{1_{1}}^{\text{(1)}}$ (and the canonical integral $\vec{W}_{1_1}$) does not appear there either.}  Furthermore, thanks to the $t\leftrightarrow u$ symmetry of both $\mathcal{A}_1^{(2)}$ and $\mathcal{A}_2^{(2)}$, we get
\begin{align}
 & c_{i;7}^{\left(n\right)}=c_{i;2}^{\left(n\right)}\Big|_{t\leftrightarrow u}\,, &  & c_{i;4}^{\left(n\right)}=-2c_{i;2}^{\left(n\right)}\Big|_{t\leftrightarrow m_{H}^{2}}\,.
\end{align}
An analogous behaviour is observed for higher transcendental weight contributions.

To ensure an efficient numerical evaluation up to $\mathcal{O}\left(\epsilon^{2}\right)$,
we follow two complementary strategies. On the one hand, we numerically evaluate the single Chen iterated integrals, according to the discussion presented in Appendix~\ref{app:cii}.
On the other hand, we solve the differential equations for the independent functions $w_{i_{k}}^{(k')}$ by means of generalised series expansion.

We present, in the supplemental material of this paper, the analytic expressions for all the non-vanishing form factors in terms of the independent functions $w_{i_{k}}^{(k')}$, along with their representation as Chen iterated integrals.

\section{Results and Checks}
\label{sec:results}

Thanks to the correspondence between the form factors $\mathcal{F}_{1,2}$ and the helicity amplitudes outlined in Eq.~\eqref{eq:ffha}, we can directly write the partonic differential cross section for $gg \to HH$ as
\begin{align}
\frac{\partial\,\hat{\sigma}}{\partial\,t}
 & =\frac{1}{2^{9}\pi s^2}\left(|F_{1}|^{2}+|F_{2}|^{2}\right),
 \label{eq:dsigma_dt}
\end{align}
with $F_i = \mathcal{F}_i + \mathcal{F}_{\textup{LO},i}\,,$
where $\mathcal{F}_i$ are defined as in Eq.~\eqref{eq:ffs} and the $\mathcal{F}_{\textup{LO},i}$ are obtained applying the projectors of Eq.~\eqref{eq:proj} to the one-loop, top-mediated $gg \to HH$ amplitude~\cite{Glover:1987nx,Plehn:1996wb}.

The phase-space for the production of a Higgs boson pair reads
\begin{align}
 & s>4m_H^{2} &  & 
 \& &&
 m_H^{2}-\frac{s}{2}\left(1+\beta_H\right)<t<m_H^{2}-\frac{s}{2}\left(1-\beta_H\right)\,,
 \label{eq:phy_reg_st}
\end{align}
with $\beta_H^{2}=1-4m_H^2/s$. We can also parametrise the process in terms
of the dimensionless variables $\eta$ and $\phi$, defined as
\begin{align}
 & \eta=\frac{s}{4m_H^{2}}-1\,, &  & \phi=\frac{m_H^{2}-t}{s}\,.
 \label{eq:etaphi_def}
\end{align}
Using these variables, the physical region described in Eq.~\eqref{eq:phy_reg_st} becomes
\begin{align}
 & \eta>0 &  & \& &  & \frac{1}{2}\left(1-\sqrt{\frac{\eta}{1+\eta}}\right)<\phi<\frac{1}{2}\left(1+\sqrt{\frac{\eta}{1+\eta}}\right)\,.
 \label{eq:etaphi}
\end{align}

We now aim to numerically evaluate the functions $\mathcal{A}_1^{(2)}$, $\mathcal{A}_2^{(2)}$, and $\mathcal{A}_3^{(2)}$ from Sec.~\ref{sec:amplitude}. 
Specifically, we are interested in analytically continuing our independent functions (as well as our canonical integrals) from the region $0<s<m_V^2\,,u,t<0$ to the production region of Eq.~\eqref{eq:phy_reg_st}. 

Let us begin this analysis by numerically comparing our results for $\mathcal{A}_3^{(2)}$ against the analytic expressions, in terms of generalised polylogarithms, reported in~\cite{Bonetti:2016brm}. 
To perform this analysis, we solve the differential equations for all independent functions $w_{i_k}^{(k')}$ present in our analytic expressions for $A_3^{(2)}$ up to $\mathcal{O}\left(\epsilon^2\right)$, as explained in Sec.~\ref{secsub:ffs}.

Notice that, to reach the production region evolving from the large-mass limit,
we have to cross the thresholds
\begin{align}
 & m_{V}^{2}-s\,, &  & 4m_{V}^{2}-s\,, &  & m_{V}^{2}-m_H^{2}\,, &  & m_H^{2}-s\,, &  & 4m_H^{2}-s\,.
 \label{eq:threshold}
\end{align}
\begin{figure}[t]
\centering
\subfloat[$\mathcal{O}\left(\epsilon^{0}\right)$]{\includegraphics[scale=0.75]{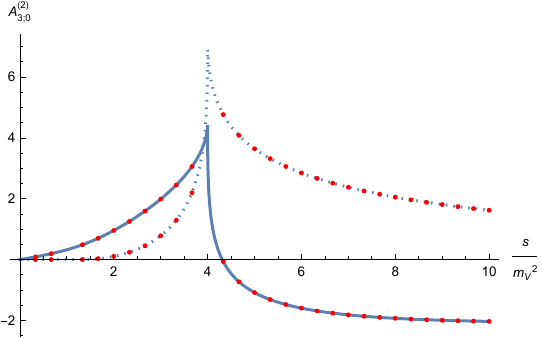}}\qquad
\subfloat[$\mathcal{O}\left(\epsilon^{1}\right)$]{\includegraphics[scale=0.75]{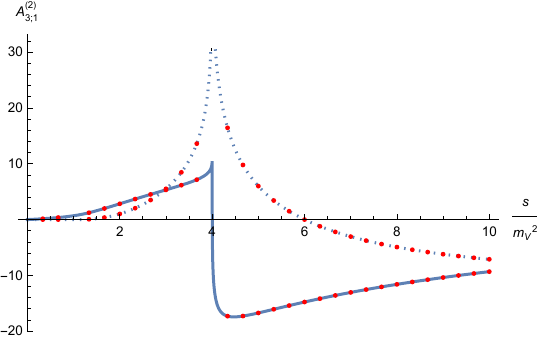}}\qquad
\subfloat[$\mathcal{O}\left(\epsilon^{2}\right)$]{\includegraphics[scale=0.75]{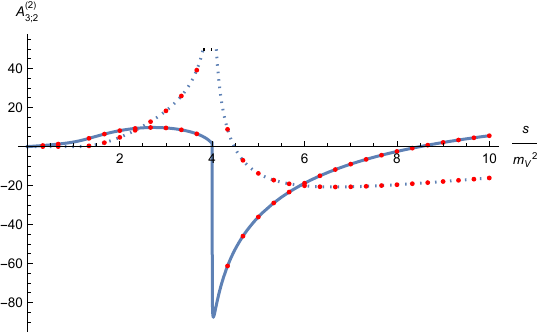}}
\caption{Two-dimensional plots for real (solid line) and imaginary (dashed line) 
parts of $\mathcal{A}_3^{(2)}$ as a function of $s/m_V^2$, 
up to $\mathcal{O}\left(\epsilon^2\right)$ generated by numerical evaluation of $w_{i_k}^{(k')}$ via generalised power series expansion through \textsc{DiffExp} (blue) and numerical evaluation of analytic expressions in terms of generalised polylogarithms (red). 
} 
\label{fig:plot_A3}
\end{figure}
For the evaluation of $\mathcal{A}_3^{(2)}$, we only need to cross the first and second thresholds, since this form factor depends only on $s$ and $m_V^2$. 
We solve the differential equations and perform the required analytic continuations with the help of {\sc DiffExp}, by  giving a small imaginary part to $s$ ($s\to s+\iu\delta$). 

In Fig.~\ref{fig:plot_A3}, we show the comparison between our numerical evaluation via generalised power series expansion and the numerical evaluation of generalised polylogarithms.
For the comparison, we evolve along the straight line 
\begin{align}
    \vec{x} = \vec{x}_0\,(1-y) + \vec{x}_1\, y\,,
\label{eq:line}
\end{align}
with $\vec{x}_0=\{0,0,0,1\}$ and 
$\vec{x}_{1} = \{10,0,0,1\}$, 
and $y\in[0,1]$.
Through this path, {\sc DiffExp} provides generalised power series expansions that allow for fast numerical evaluations. 

~

Let us now consider $\mathcal{A}_1^{(2)}$ and $\mathcal{A}_2^{(2)}$.
We first transport
our boundary values to a point in the production
region $s>4m_H^{2}$, and then use this new point as a base point to explore the whole physical region. This allows us to skip performing the necessary analytic continuations every time we evolve to a new point from the large-mass limit and cross the thresholds listed in Eq.~\eqref{eq:threshold}. 
Without loss of generality, we pick the phase-space point
\begin{align}
\left\{ s_{0}\,,t_{0}\,,u_{0}\,,m_{H;0}^{2}\,,m_{V;0}^{2}\right\}  & =\left\{ \frac{3125}{256}\,,-\frac{1875}{512}\,,-\frac{1875}{512}\,,\frac{625}{256}\,,1\right\} \,.
\label{eq:psp_test}
\end{align}
It is worth mentioning that the numerical evaluation of our independent functions required constructing a system of 220 differential equations. To make our work self-contained, we provide a {\sc Mathematica} notebook in the ancillary files to numerically solve these differential equations using {\sc DiffExp}.
Furthermore, to ensure that the analytic continuation has been correctly performed, we evaluate the canonical integrals at this kinematic point using {\sc AMFlow}~\cite{Liu:2017jxz,Liu:2022chg}, finding agreement.
In Table~\ref{tab:FFs_num_eval}, we present numerical values for the form factors $\mathcal{A}_1^{(2)}$, $\mathcal{A}_2^{(2)}$, and $\mathcal{A}_3^{(2)}$ at this phase-space point. 

\begin{table}[t]
\centering
\small
    \begin{tabular}{cccc}
        \toprule
        {} & $\mathcal{A}_1^{(2)}$ & $\mathcal{A}_2^{(2)}$ & $\mathcal{A}_3^{(2)}$\\
        \midrule
        \multirow{2}{*}{$\epsilon^0$} & 
        $7.6630892513031246689$ & $0.32099028648379086709$ & $-2.0639052901232038861$ \\
        & $+ 1.2522181718007465702\iu$ & $+ 0.52693547002550936565\iu$ & $+1.3199524961270220223\iu$
        \\[0.1cm]
        \multirow{2}{*}{$\epsilon^1$} &
        $11.583682591428381615$ & $-1.6327683705293072671$ & $-7.4642615923306829442$ \\
        & $+ 22.667681416494406992 \iu$ & $+ 1.7276827046764600459\iu$ & $- 8.4210262675049745787 \iu$
        \\[0.1cm]
        \multirow{2}{*}{$\epsilon^2$} & $-12.799667397909851909$ & $-5.0540312586637065786$ & $10.697978848021963295$ \\
        & $+ 33.636555225786291772\iu$ & $- 0.6727314555148337712\iu$ & $- 12.068177949009557105\iu$
        \\
      \bottomrule
    \end{tabular}
\caption{
Real and imaginary parts of the numerical evaluation of the functions $\mathcal{A}_1^{(2)}$, $\mathcal{A}_2^{(2)}$, and $\mathcal{A}_3^{(2)}$ in the phase-space point of Eq.~\eqref{eq:psp_test}.
}\label{tab:FFs_num_eval}
\end{table}

\begin{figure}[t]
\centering
\subfloat[$\mathcal{O}\left(\epsilon^0\right)$]{
\includegraphics[scale=0.75]{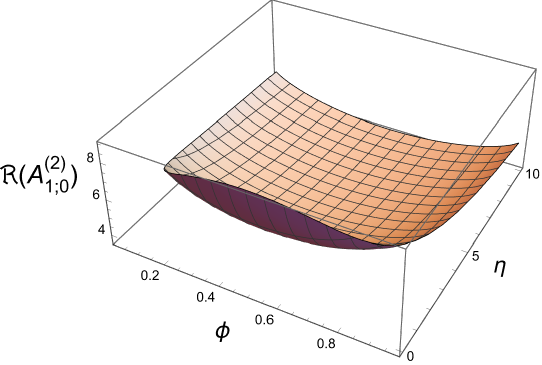}\qquad
\includegraphics[scale=0.75]{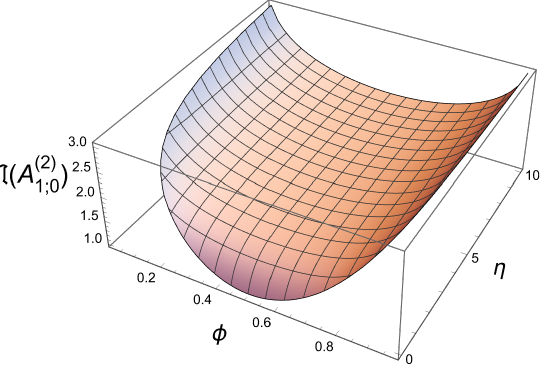}
}\qquad\\
\subfloat[$\mathcal{O}\left(\epsilon^1\right)$]{
\includegraphics[scale=0.75]{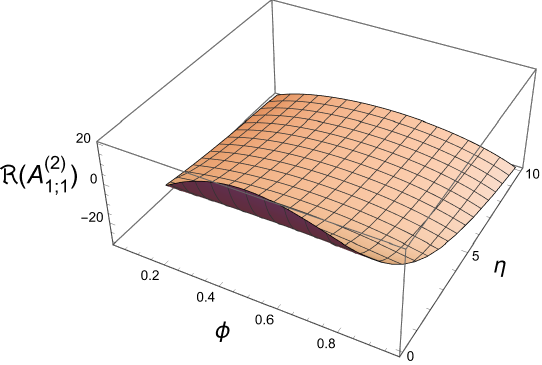}\qquad
\includegraphics[scale=0.75]{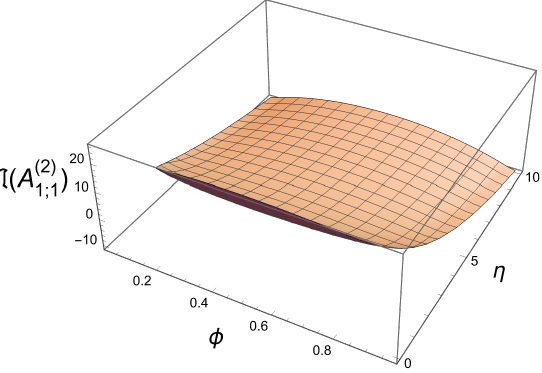}
}\qquad\\
\subfloat[$\mathcal{O}\left(\epsilon^2\right)$]{
\includegraphics[scale=0.75]{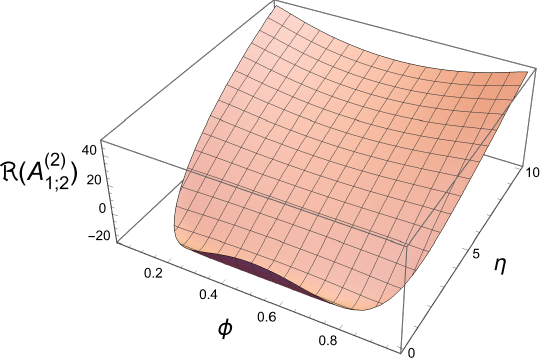}\qquad
\includegraphics[scale=0.75]{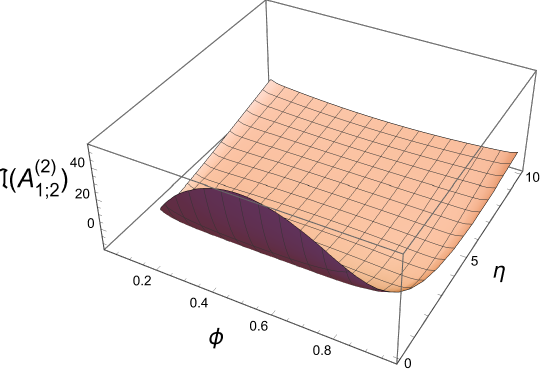}
}
 \caption{
 Three-dimensional plots of the real and imaginary parts of $\mathcal{A}_1^{(2)}$ as a function of $\eta$ and $\phi$ 
 in the physical region of Eq.~\eqref{eq:etaphi}. 
} 
\label{fig:plot_A1}
\end{figure}

\begin{figure}[t]
\centering
\subfloat[$\mathcal{O}\left(\epsilon^0\right)$]{
\includegraphics[scale=0.75]{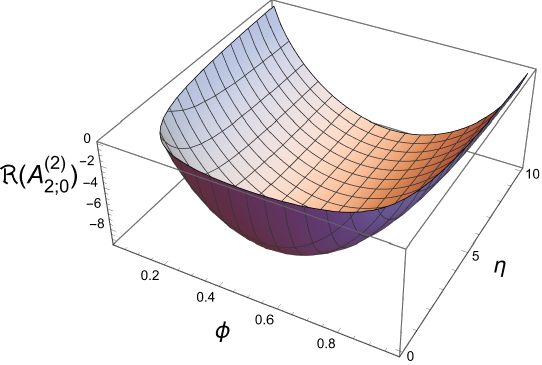}\qquad
\includegraphics[scale=0.75]{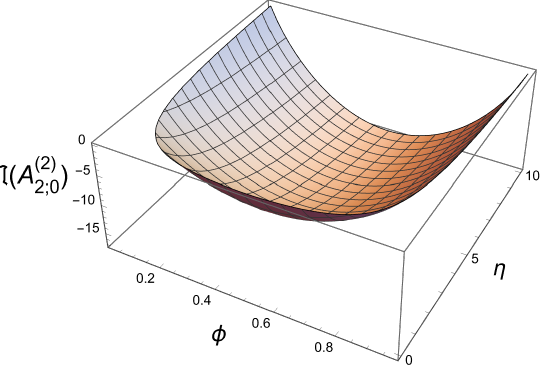}
}\qquad
\subfloat[$\mathcal{O}\left(\epsilon^1\right)$]{
\includegraphics[scale=0.75]{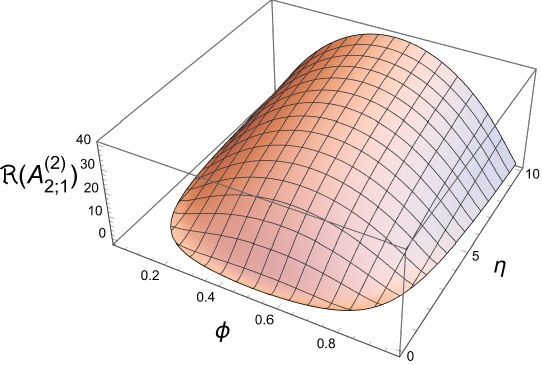}\qquad
\includegraphics[scale=0.75]{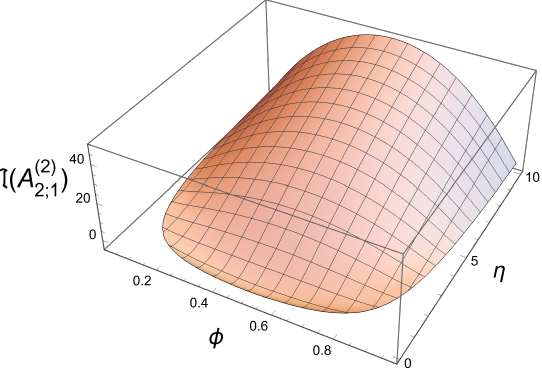}
}\qquad
\subfloat[$\mathcal{O}\left(\epsilon^2\right)$]{
\includegraphics[scale=0.75]{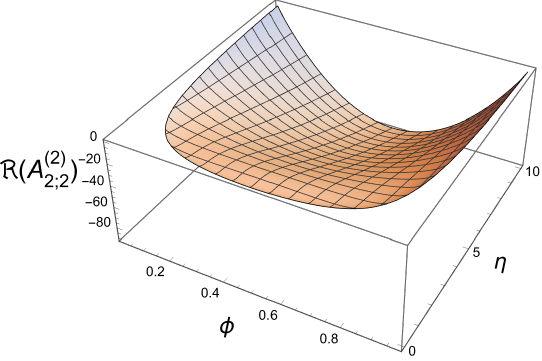}\qquad
\includegraphics[scale=0.75]{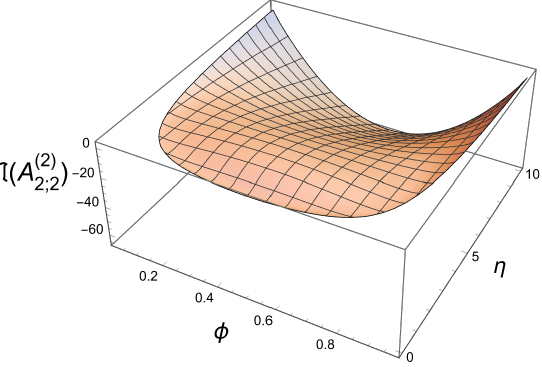}
}
 \caption{
 Three-dimensional plots of the real and imaginary parts of $\mathcal{A}_2^{(2)}$ as a function of $\eta$ and $\phi$ 
 in the physical region of Eq.~\eqref{eq:etaphi}.
} 
\label{fig:plot_A2}
\end{figure}

We demonstrate the efficiency and reliability of our setup for numerical evaluations
by focusing on $\mathcal{A}_1^{(2)}$ and $\mathcal{A}_2^{(2)}$ from $\epsilon^0$, through $\epsilon^2$. We construct a grid of 4,143 points in terms of $\eta$ and $\phi$ variables in the production region, which we use to produce the three-dimensional plots presented in Figs.~\ref{fig:plot_A1} and~\ref{fig:plot_A2}. The first 1,600 points are chosen to be same as in~\cite{Czakon:2008zk,Mandal:2022vju}, whereas the remaining 
2,543 are distributed on an evenly spaced grid near the threshold $s = 4 m_H^2$. 
We experienced a CPU time required per phase-space point  of $\mathcal{O}(5')$ (starting from the large mass expansion) and $\mathcal{O}(40'')$ (starting from a point in the physical region), on a desktop machine with processor {\sc AMD RYZEN 9 9700X} and 32 GB of DDR5 RAM.
These grids are available from the authors upon request.

Let us emphasise that, due to the organisation of our amplitude in terms of independent functions, we avoid large cancellations and spurious poles in intermediate steps of the computation. 
As a result, the accuracy of the amplitude is determined mainly by the numerical precision achieved in the evaluation of the independent functions. 
For a technical discussion on the computation, see Appendix~\ref{app:computation}.

\clearpage
\section{Conclusions}
\label{sec:conclusions}

In this paper, we computed analytic expressions for the electroweak (EW) two-loop, light-quark contributions to the two form factors that describe double Higgs boson production via gluon fusion.
Since this class of contributions appear for the first time at two-loops,
we employed dimensional regularisation and standard methods for calculating multi-loop scattering amplitudes, observing explicit cancellations of infrared and ultraviolet  divergences.

To evaluate these form factors, we identified three independent gauge-invariant groups: two of them containing three-point Feynman integrals and one containing genuine four-point ones. While the three-point integrals are already known, the four-point integrals required explicit calculation, revealing their dependence on four kinematic invariants. We computed these integrals using the method of differential equations in canonical form, constructing an independent basis with uniform transcendental weight and solving them in terms of Chen iterated integrals.

For fast numerical evaluation of the form factors,  we expressed our results in terms of generalised power series expansions with the aid of the {\sc Mathematica} package {\sc DiffExp}. We elaborated on a procedure to evaluate the linear combinations of Feynman integrals appearing in the amplitude in terms of independent functions by constructing a system of differential equations that is independent of the dimensional regulator $\epsilon$, avoiding the computation of unnecessary terms.
This procedure led to remarkable simplifications in the finite contribution to the form factors while highlighting the increasing complexity of higher orders in $\epsilon$, which are essential for next-to-leading order calculations.

Our results open several future research directions:
\begin{enumerate}
    \item The analytic expressions of the form factors provided in this paper, supplemented with dedicated tools for their numerical evaluation, are ready for implementation into codes for phenomenological studies, such as the \textsc{ggHH} library \cite{Heinrich:2017kxx,Heinrich:2019bkc,Heinrich:2020ckp} of the \textsc{POWHEG-BOX-V2} \cite{Nason:2004rx,Frixione:2007vw,Alioli:2010xd}, which will be the topic of a dedicated publication.
    This will allow to investigate the impact of light-quark EW corrections on several observables, and represents a major step towards more precise studies of double Higgs boson production at hadron colliders.
    
    \item The analytic expressions for the two-loop Feynman integrals calculated in this work, as well as the computational framework and organisation of amplitudes in terms of uniform weight functions, can be adapted to other scattering processes that manifest similar kinematic configuration, such as higher-order vector boson pair production and Higgs plus vector boson production. The techniques described here can be implemented in computer codes for efficient evaluation of scattering amplitudes through the solution of differential equations, such as \textsc{SeaSyde}~\cite{Armadillo:2022ugh} or \textsc{Line}~\cite{Prisco:2025wqs}.
    
    \item NLO QCD corrections have been shown to increase the cross section for double Higgs production in gluon fusion by $\mathcal{O}\left(+60\%\right)$ w.r.t.\ LO~\cite{Borowka:2016ehy,Borowka:2016ypz,Baglio:2018lrj,Davies:2019dfy,Baglio:2020ini}. Aiming at a percent-level theoretical uncertainty, QCD corrections applied to light-quark EW contributions cannot be discarded and will require dedicated work. Even though we expect an increase in complexity when addressing such contributions, we are convinced that the space of functions describing the Feynman integrals and the amplitude will not change, allowing us to apply the results and methods outlined here to the computation of mixed QCD-EW corrections to double Higgs production at three loops.
\end{enumerate}
\section*{Acknowledgments}

We wish to thank Gudrun Heinrich and Matthias Kerner for support and collaboration during the early stages of this project, and Antonela Matijašić and Julian Miczajka for discussions and sharing their package {\sc Effortless} with us prior to publication.
M.B.\ and W.J.T.\ acknowledge the Mainz Institute for Theoretical Physics (MITP) of the Cluster of Excellence $\text{PRISMA}^+$ (Project ID 390831469), for its hospitality and support during the workshop \textit{The Evaluation of the Leading Hadronic Contribution to the Muon $g-2$: Consolidation of the MUonE Experiment and Recent Developments in Low Energy $e^+$ $e^-$ Data}.
This work is supported by the \textit{Deutsche Forschungsgemeinschaft} (DFG, German Research Foundation) under grant no.\ 396021762 - TRR 257, and by the Leverhulme Trust, LIP-2021-01.

\appendix

 \section{The alphabet}
\label{app:alphabet}

We present the complete alphabet of the canonical differential equations~\eqref{eq:deq_dlog}. We split it into even and algebraic letters.

\subsubsection*{Even letters}

\begin{align}
 & \alpha_{0}=m_{V}^{2}\,, &  &  \\
 & \alpha_{1}=m_H^{2}\,, &  & \alpha_{16}=4m_{V}^{2}\left(m_{H}^{4}-tu\right)-st^{2}\,,\nonumber \\
 & \alpha_{2}=s\,, &  & \alpha_{17}=m_H^{4}-m_{V}^{2}\left(4m_H^{2}-s\right)\,,\nonumber \\
 & \alpha_{3}=t\,, &  & \alpha_{18}=\left(m_H^{2}-m_{V}^{2}\right)^{2}+sm_{V}^{2}\,,\nonumber \\
 & \alpha_{4}=u\,, &  & \alpha_{19}=m_H^{4}-tu\,,\nonumber \\
 & \alpha_{5}=4m_{V}^{2}-s\,, &  & \alpha_{20}=m_{V}^{2}(t-u)^{2}+stu\,,\nonumber \\
 & \alpha_{6}=4m_{V}^{2}-m_H^{2}\,, &  & \alpha_{21}=m_H^{2}\left(m_{V}^{2}-t\right)+t^{2}\,,\nonumber \\
 & \alpha_{7}=4m_H^{2}-s\,, &  & \alpha_{22}=m_H^{2}\left(m_{V}^{2}-u\right)+u^{2}\,,\nonumber \\
 & \alpha_{8}=m_{V}^{2}-m_H^{2}\,, &  & \alpha_{23}=m_{V}^{2}\left(m_H^{2}-u\right)-st\,,\nonumber \\
 & \alpha_{9}=m_{V}^{2}-s\,, &  & \alpha_{24}=m_{V}^{2}\left(m_H^{2}-t\right)-su\,,\nonumber \\
 & \alpha_{10}=m_{V}^{2}-t\,, &  & \alpha_{25}=m_H^{4}-sm_{V}^{2}-tu\,,\nonumber \\
 & \alpha_{11}=m_{V}^{2}-u\,, &  & \alpha_{26}=4m_{V}^{2}\left(sm_H^{2}+tu-m_H^{4}\right)-m_H^{4}s\,,\nonumber \\
 & \alpha_{12}=m_H^{2}-t\,, &  & \alpha_{27}=sm_H^{2}\left(m_H^{2}-u\right)+m_{V}^{2}\left(m_H^{2}-t\right)^{2}\,,\nonumber \\
 & \alpha_{13}=m_H^{2}-u\,, &  & \alpha_{28}=sm_H^{2}\left(m_H^{2}-t\right)+m_{V}^{2}\left(m_H^{2}-u\right)^{2}\,,\nonumber \\
 & \alpha_{14}=m_{V}^{2}\left(s^{2}+m_{V}^{2}\left(m_{V}^{2}-3s\right)\right)+sm_H^{4}\,, &  & \alpha_{29}=m_{V}^{4}\left(4m_H^{2}-s\right)+2tm_{V}^{2}(t-u)-st^{2}\,,\nonumber \\
 & \alpha_{15}=4m_{V}^{2}\left(m_{H}^{4}-tu\right)-su^{2}\,, &  & \alpha_{30}=m_{V}^{4}\left(4m_H^{2}-s\right)+2um_{V}^{2}(u-t)-su^{2}\,.\nonumber 
\end{align}

\subsubsection*{Algebraic letters}

\begin{itemize}
    \item Odd parity letters of first type
\begin{align}
 & \alpha_{31}=\frac{t-u-r_{1}}{t-u+r_{1}}\,,\\
 & \alpha_{32}=\frac{2m_{V}^{2}-2m_H^{2}+s-r_{1}}{2m_{V}^{2}-2m_H^{2}+s+r_{1}}\,, &  & \alpha_{41}=\frac{m_H^{2}-2m_{V}^{2}-r_{5}}{m_H^{2}-2m_{V}^{2}+r_{5}}\,,\nonumber \\
 & \alpha_{33}=\frac{2m_H^{2}-s-r_{1}}{2m_H^{2}-s+r_{1}}\,, &  & \alpha_{42}=\frac{2m_H^{2}m_{V}^{2}-2tm_{V}^{2}-su-r_{6}}{2m_H^{2}m_{V}^{2}-2tm_{V}^{2}-su+r_{6}}\,,\nonumber \\
 & \alpha_{34}=\frac{2m_H^{2}m_{V}^{2}-2um_{V}^{2}-st-r_{2}}{2m_H^{2}m_{V}^{2}-2um_{V}^{2}-st+r_{2}}\,, &  & \alpha_{43}=\frac{su-r_{6}}{su+r_{6}}\,,\nonumber \\
 & \alpha_{35}=\frac{st-r_{2}}{st+r_{2}}\,, &  & \alpha_{44}=\frac{s\left(2m_{V}^{2}-m_H^{2}\right)-r_{7}}{s\left(2m_{V}^{2}-m_H^{2}\right)+r_{7}}\,,\nonumber \\
 & \alpha_{36}=\frac{(t-u)m_{V}^{2}-st-r_{3}}{(t-u)m_{V}^{2}-st+r_{3}}\,, &  & \alpha_{45}=\frac{2m_{V}^{2}m_H^{2}-sm_H^{2}-2um_{V}^{2}-r_{7}}{2m_{V}^{2}m_H^{2}-sm_H^{2}-2um_{V}^{2}+r_{7}}\,,\nonumber \\
 & \alpha_{37}=\frac{s\left(m_{V}^{2}-t\right)-r_{3}}{s\left(m_{V}^{2}-t\right)+r_{3}}\,, &  & \alpha_{46}=\frac{2m_{V}^{2}m_H^{2}-sm_H^{2}-2tm_{V}^{2}-r_{7}}{2m_{V}^{2}m_H^{2}-sm_H^{2}-2tm_{V}^{2}+r_{7}}\,,\nonumber \\
 & \alpha_{38}=\frac{2m_{V}^{2}-s-r_{4}}{2m_{V}^{2}-s+r_{4}}\,, &  & \alpha_{47}=\frac{2m_H^{4}-sm_H^{2}-2tu-r_{7}}{2m_H^{4}-sm_H^{2}-2tu+r_{7}}\,,\nonumber \\
 & \alpha_{39}=\frac{m_H^{2}-2t-r_{5}}{m_H^{2}-2t+r_{5}}\,, &  & \alpha_{48}=\frac{(t-u)m_{V}^{2}+su-r_{8}}{(t-u)m_{V}^{2}+su+r_{8}}\,,\nonumber \\
 & \alpha_{40}=\frac{m_H^{2}-2u-r_{5}}{m_H^{2}-2u+r_{5}}\,, &  & \alpha_{49}=\frac{s\left(m_{V}^{2}-u\right)-r_{8}}{s\left(m_{V}^{2}-u\right)+r_{8}}\,.\nonumber 
\end{align}
    \item Mixed parity letters
\begin{align}
 & \alpha_{50}=\frac{s\left(2m_H^{4}-2tm_H^{2}+st\right)-r_{1}r_{2}}{s\left(2m_H^{4}-2tm_H^{2}+st\right)+r_{1}r_{2}}\,, \\
 & \alpha_{51}=\frac{s\left(2m_H^{4}-2\left(2m_{V}^{2}+t\right)m_H^{2}+s\left(m_{V}^{2}+t\right)\right)-r_{1}r_{3}}{s\left(2m_H^{4}-2\left(2m_{V}^{2}+t\right)m_H^{2}+s\left(m_{V}^{2}+t\right)\right)+r_{1}r_{3}}\,,\nonumber \\
 & \alpha_{52}=\frac{s\left(2m_H^{2}-s\right)-r_{1}r_{4}}{s\left(2m_H^{2}-s\right)+r_{1}r_{4}}\,,\nonumber \\
 & \alpha_{53}=\frac{m_H^{2}\left(2m_{V}^{2}+s\right)-r_{1}r_{5}}{\left(2m_{V}^{2}+s\right)m_H^{2}+r_{1}r_{5}}\,,\nonumber \\
 & \alpha_{54}=\frac{-2m_H^{4}+\left(8m_{V}^{2}+s\right)m_H^{2}-2sm_{V}^{2}-r_{1}r_{5}}{-2m_H^{4}+\left(8m_{V}^{2}+s\right)m_H^{2}-2sm_{V}^{2}+r_{1}r_{5}}\,,\nonumber \\
 & \alpha_{55}=\frac{s\left(2m_H^{4}-2um_H^{2}+su\right)-r_{1}r_{6}}{s\left(2m_H^{4}-2um_H^{2}+su\right)+r_{1}r_{6}}\,,\nonumber \\
 & \alpha_{56}=\frac{s\left(2m_H^{4}-2\left(2m_{V}^{2}+u\right)m_H^{2}+s\left(m_{V}^{2}+u\right)\right)-r_{1}r_{8}}{s\left(2m_H^{4}-2\left(2m_{V}^{2}+u\right)m_H^{2}+s\left(m_{V}^{2}+u\right)\right)+r_{1}r_{8}}\,,\nonumber \\
 & \alpha_{57}=\frac{2sm_{V}^{2}m_H^{4}+st\left((t-3u)m_{V}^{2}-st\right)-r_{2}r_{3}}{2sm_{V}^{2}m_H^{4}+st\left((t-3u)m_{V}^{2}-st\right)+r_{2}r_{3}}\,,\nonumber \\
 & \alpha_{58}=\frac{-s\left(2(u-t)m_{V}^{2}+st\right)-r_{2}r_{4}}{-s\left(2(u-t)m_{V}^{2}+st\right)+r_{2}r_{4}}\,,\nonumber \\
 & \alpha_{59}=\frac{2\left(m_H^{2}+s\right)m_{V}^{2}m_H^{2}-stm_H^{2}-2tum_{V}^{2}-r_{2}r_{5}}{2\left(m_H^{2}+s\right)m_{V}^{2}m_H^{2}-stm_H^{2}-2tum_{V}^{2}+r_{2}r_{5}}\,,\nonumber \\
 & \alpha_{60}=\frac{s\left(2m_H^{4}-\left(2m_{V}^{2}+t\right)m_H^{2}-2um_{V}^{2}\right)-r_{2}r_{5}}{s\left(2m_H^{4}-\left(2m_{V}^{2}+t\right)m_H^{2}-2um_{V}^{2}\right)+r_{2}r_{5}}\,,\nonumber \\
 & \alpha_{61}=\frac{-s\left((u-3t)m_{V}^{2}+st\right)-r_{3}r_{4}}{-s\left((u-3t)m_{V}^{2}+st\right)+r_{3}r_{4}}\,,\nonumber \\
 & \alpha_{62}=\frac{s\left(\left(m_{V}^{2}-t\right)m_H^{2}+2tm_{V}^{2}\right)-r_{3}r_{5}}{s\left(\left(m_{V}^{2}-t\right)m_H^{2}+2tm_{V}^{2}\right)+r_{3}r_{5}}\,,\nonumber \\
 & \alpha_{63}=\frac{s\left(m_H^{2}-2m_{V}^{2}\right)-r_{4}r_{5}}{s\left(m_H^{2}-2m_{V}^{2}\right)+r_{4}r_{5}}\,,\nonumber \\
 & \alpha_{64}=\frac{s\left(2(t-u)m_{V}^{2}+su\right)-r_{4}r_{6}}{s\left(2(t-u)m_{V}^{2}+su\right)+r_{4}r_{6}}\,,\nonumber \\
 & \alpha_{65}=\frac{-\left(\left(m_{H}^{2}-2m_{V}^{2}\right)s^{2}\right)-r_{4}r_{7}}{-s^{2}\left(m_{H}^{2}-2m_{V}^{2}\right)+r_{4}r_{7}}\,,\nonumber \\
 & \alpha_{66}=\frac{s\left((t-3u)m_{V}^{2}+su\right)-r_{4}r_{8}}{s\left((t-3u)m_{V}^{2}+su\right)+r_{4}r_{8}}\,,\nonumber \\
 & \alpha_{67}=\frac{2m_H^{2}\left(m_H^{2}+s\right)m_{V}^{2}-u\left(sm_H^{2}+2tm_{V}^{2}\right)-r_{5}r_{6}}{2m_H^{2}\left(m_H^{2}+s\right)m_{V}^{2}-u\left(sm_H^{2}+2tm_{V}^{2}\right)+r_{5}r_{6}}\,,\nonumber \\
 & \alpha_{68}=\frac{s\left(2m_H^{4}-\left(2m_{V}^{2}+u\right)m_H^{2}-2tm_{V}^{2}\right)-r_{5}r_{6}}{s\left(2m_H^{4}-\left(2m_{V}^{2}+u\right)m_H^{2}-2tm_{V}^{2}\right)+r_{5}r_{6}}\,,\nonumber \\
 & \alpha_{69}=\frac{-\left(\left(2m_{V}^{2}+s\right)m_H^{4}\right)+4sm_{V}^{2}m_H^{2}+2tum_{V}^{2}-r_{5}r_{7}}{-\left(\left(2m_{V}^{2}+s\right)m_H^{4}\right)+4sm_{V}^{2}m_H^{2}+2tum_{V}^{2}+r_{5}r_{7}}\,,\nonumber \\
 & \alpha_{70}=\frac{s\left(\left(m_{V}^{2}-u\right)m_H^{2}+2um_{V}^{2}\right)-r_{5}r_{8}}{s\left(\left(m_{V}^{2}-u\right)m_H^{2}+2um_{V}^{2}\right)+r_{5}r_{8}}\,,\nonumber \\
 & \alpha_{71}=\frac{2sm_{V}^{2}m_H^{4}+su\left((u-3t)m_{V}^{2}-su\right)-r_{6}r_{8}}{2sm_{V}^{2}m_H^{4}+su\left((u-3t)m_{V}^{2}-su\right)+r_{6}r_{8}}\nonumber \,.
\end{align}
    \item Odd parity letters of second type
\begin{align}
 & \alpha_{72}=\frac{-m_H^{4}+\left(u-2m_{V}^{2}\right)m_H^{2}+2tm_{V}^{2}+\left(m_H^{2}-u\right)r_{5}}{m_H^{4}+2m_{V}^{2}m_H^{2}-um_H^{2}-2tm_{V}^{2}+\left(m_H^{2}-u\right)r_{5}}\,, \\
 & \alpha_{73}=\frac{\left(-3m_H^{2}+2t+u\right)m_H^{2}+\left(m_H^{2}-u\right)r_{5}}{\left(3m_H^{2}-2t-u\right)m_H^{2}+\left(m_H^{2}-u\right)r_{5}}\,,\nonumber \\
 & \alpha_{74}=\frac{-m_H^{4}+3m_{V}^{2}m_H^{2}-2sm_{V}^{2}+\left(m_H^{2}-m_{V}^{2}\right)r_{5}}{m_H^{4}-3m_{V}^{2}m_H^{2}+2sm_{V}^{2}+\left(m_H^{2}-m_{V}^{2}\right)r_{5}}\,,\nonumber \\
 & \alpha_{75}=\frac{-s^{2}+3m_H^{2}s-2m_H^{2}m_{V}^{2}+\left(m_H^{2}-s\right)r_{1}}{-s^{2}+3m_H^{2}s-2m_H^{2}m_{V}^{2}+\left(s-m_H^{2}\right)r_{1}}\,,\nonumber \\
 & \alpha_{76}=\frac{2(u-t)m_{V}^{2}-su-ur_{4}}{2(u-t)m_{V}^{2}-su+ur_{4}}\,,\nonumber \\
 & \alpha_{77}=\frac{-\left(\left(2(t-u)m_{V}^{2}+su\right)m_H^{2}\right)+2(t-u)um_{V}^{2}-ur_{7}}{-\left(\left(2(t-u)m_{V}^{2}+su\right)m_H^{2}\right)+2(t-u)um_{V}^{2}+ur_{7}}\nonumber \,.
\end{align}
\end{itemize}
 \section{Details on the large-mass expansion}
\label{app:LME}

The large-mass expansion corresponds to the limit $s,t,u,m_H^2 \ll m_V^2$. The induced hierarchy of scales sets all integral regions to zero except for those where the momentum flowing in the propagators is either \emph{large} ($k^2 \sim m_V^2$) or \emph{small} ($k^2 \sim s,t,u,m_H^2$). In particular, momentum conservation implies that large momentum must be flowing in closed loops within the Feynman integral, since it cannot be provided by the external legs. In practical terms, we need to identify all possible inequivalent large-momentum configurations and parametrise them in terms of a set of independent large and small loop momenta. We then proceed to expand around the large momenta in terms of the small ones.

To illustrate the procedure, consider the two-loop planar 
and non-planar integrals depicted in Fig.~\ref{fig:tsecGG}, corresponding to the top sectors for our study.
We identify five regions of interest, which, in terms of the loop momenta listed in Table~\ref{tab:toposGG}, correspond to the following configurations:
\begin{itemize}
\begin{subequations}
\item Both loop momenta are small:
\begin{align}
k_1^2, k_2^2 \ll m_V^2\,,\label{eq:reg1}
\end{align}
with $ (k_1 - k_2)^2 \ll m_V^2$.
\item One loop momentum is large and the other one is small:
    \begin{align}
        k_1^2 &\ll m_V^2 \sim k_2^2\,,\label{eq:reg2}\\
        k_2^2 &\ll m_V^2 \sim k_1^2\,.\label{eq:reg3}
    \end{align}
In both cases, the difference satisfies $ (k_1 - k_2)^2 \sim m_V^2 $.
\item Both loop momenta are large: $ k_1^2 \sim k_2^2 \sim m_V^2 $. Here, we further distinguish between their difference being large or small:  
    \begin{align}
        (k_1 - k_2)^2 &\sim m_V^2\,,\label{eq:reg4} \\
        (k_1 - k_2)^2 &\ll m_V^2\,.\label{eq:reg5}
    \end{align}
\end{subequations}
\end{itemize}

In the planar case, we get scaleless integrals for the regions~\eqref{eq:reg1},~\eqref{eq:reg3} and~\eqref{eq:reg5}.
As a result, the only non-zero regions are~\eqref{eq:reg2} and~\eqref{eq:reg4}.

The non-zero regions can graphically be represented as
\begin{align}
    \left.\vcenter{\hbox{\includegraphics{Images/LME/PLbox.pdf}}}\right.
        &\;\to\;
    \left.\vcenter{\hbox{\includegraphics{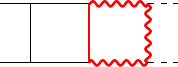}}}\right.
    \;+\;
    \left.\vcenter{\hbox{\includegraphics{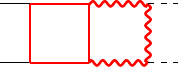}}}\right.\,,
\label{eq:PL_LME_example}
\end{align}%
where thick red lines indicate a large momentum flow. At leading order in the large-mass expansion of the integrands, the propagators with a large momentum decouple from the rest of the diagram, giving
\begin{align}
    \left.\vcenter{\hbox{\includegraphics{Images/LME/PLbox.pdf}}}\right.
    &\;\to\;
    \left.\vcenter{\hbox{\includegraphics{Images/LME/triangle.pdf}}}\right.
    \;\times\;
    \left.\vcenter{\hbox{\includegraphics{Images/LME/bubble.pdf}}}\right.
    \;+\; 
    \left.\vcenter{\hbox{\includegraphics{Images/LME/sunset1.pdf}}}\right.\,.
\label{eq:PL_app_LME}
\end{align}
We compute the resulting integrals by direct integration, obtaining
\begin{align}
    s\, \FIPL{1,1,1,1,1,1,1,0,0} \to \frac{1}{m_V^4}\left(\frac{m_V^2}{-s}\right)^\epsilon\left[\frac{1}{\epsilon^2}+\frac{2}{\epsilon}+2 + \left(2 - \frac{8\, \zeta_3}{3}\right)\epsilon+\mathcal{O}\left(\epsilon^2\right)\right].
\label{eq:PL_LME_expl}
\end{align}
Let us remark that the two-loop vacuum contribution in \eqref{eq:PL_app_LME} can be dropped 
as it is suppressed by one power of $m_V^2$ w.r.t. the remaining contributions. 
Starting from Eq.~\eqref{eq:PL_LME_expl}, we construct the large-mass expansion for the associated canonical integral, which reads
\begin{align}
    \epsilon^4\, r_3\, s\, \FIPL{ 1, 1, 1, 1, 1, 1, 1, 0, 0} &\to \frac{r_1}{m_V^2}\left(\frac{m_V^2}{-s}\right)^\epsilon\left[\epsilon^2+2 \epsilon^3+2\epsilon^4 + \left(2 - \frac{8\, \zeta_3}{3}\right)\epsilon^5+\mathcal{O}\left(\epsilon^6\right)\right]\nonumber\\
    &\to0\quad\text{for}\quad m_V^2\to\infty\,.
\end{align}
Notice that the lass mass limit has been applied to the prefactor as well (i.e. $r_3 \to m_V^2\,r_1$). 

The non-planar case is remarkably simple, as the only non-vanishing region is~\eqref{eq:reg4}.
At leading order in the expansion in $m_V^2$, this reduces to a single vacuum diagram:
\begin{align}
    \left.\vcenter{\hbox{\includegraphics{Images/LME/NPbox.pdf}}}\right.
    \;\to\; 
    \left.\vcenter{\hbox{\includegraphics{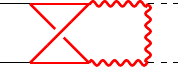}}}\right.
    \;\to\; 
    \left.\vcenter{\hbox{\includegraphics{Images/LME/sunset2.pdf}}}\right.
    \, ,
\end{align}
whose direct integration gives
\begin{small}
    \begin{align}
        m_V^6\, \FINP{1,1,1,1,1,1,1,0,0} \to -\frac{2}{\epsilon^2}-\frac{10}{\epsilon}+(-20-6\, \zeta_2)
        + \left(-24 - 30\, \zeta_2 + \frac{16\, \zeta_3}{3}\right) \epsilon +\mathcal{O}\left(\epsilon^2\right).
    \end{align}
\end{small}

By performing the large-mass expansion for the basis of canonical integrals $\vec{J}$ calculated in this work, we observe that these integrals reduce, as expected, to uniform transcendental weight combinations of multiple zeta values (The transcendental degree is assigned as $\left[\zeta_{n}\right]\to n$ and $\left[\epsilon\right]\to-1$).
 \section{Differential equations for the transcendental functions present in \texorpdfstring{$\mathcal{A}_3^{(2)}$}{A_3^(2)}}
\label{sec:appMATRIX}

We provide the differential equations for the independent functions that appear in the evaluation of $\mathcal{A}_3^{(2)}$ (cf.~Eq.~\eqref{eq:A3}), once canonical integrals $W_{i_k}$ are expressed in terms of independent functions $w_{i_k}^{(k')}$ according to Eq.~\eqref{eq:ind_wfun}.

We need 27 independent functions to construct the closed system of differential equations 
\begin{align}
\td \vec{w}_{3;2} = \td\Omega_{3;2}\,\vec{w}_{3;2}\,. 
\end{align}
The basis $\vec{w}_{3;2}$ reads
\begin{align}
    \begin{split}
        \vec{w}_{3;2} = \left\{
        w_{1_0}^{(0)},
        w_{1_0}^{(1)},
        w_{3_1}^{(1)},
        w_{5_1}^{(1)},
        w_{1_0}^{(2)},
        w_{3_1}^{(2)},
        w_{5_2}^{(2)},
        w_{5_1}^{(2)},
        w_{11_2}^{(2)},
        w_{13_2}^{(2)},
        w_{3_1}^{(3)},
        w_{5_2}^{(3)},
        w_{5_1}^{(3)},
        w_{11_2}^{(3)},
        \right.\\\left.
        w_{13_2}^{(3)},
        w_{12_3}^{(3)},
        w_{29_3}^{(3)},
        w_{30_3}^{(3)},
        w_{5_2}^{(4)},
        w_{11_2}^{(4)},
        w_{13_2}^{(4)},
        w_{12_3}^{(4)},
        w_{29_3}^{(4)},
        w_{30_3}^{(4)},
        w_{12_3}^{(5)},
        w_{29_3}^{(5)},
        w_{30_3}^{(5)}\right\}\,,
    \end{split}
\end{align}
and the matrix of coefficients $\Omega_{3;2}$ takes the form:
\begin{align}
    \Omega_{3;2} = 
    \begin{pmatrix}
        \mathbb{M}_{18\times10} &\mathbb{O}_{18\times8} &\mathbb{O}_{18\times9} \\
        \mathbb{O}_{6 \times10} &\mathbb{N}_{6 \times8} &\mathbb{O}_{6 \times9} \\
        \mathbb{O}_{3 \times10} &\mathbb{O}_{3 \times8} &\mathbb{P}_{3 \times9} \\
    \end{pmatrix}\,,
\end{align}
with
\begin{align}
    \mathbb{M}_{18\times10} = 
    \begin{small}
        \begin{pmatrix}
            0 & 0 & 0 & 0 & 0 & 0 & 0 & 0 & 0 & 0  \\
            0 & 0 & 0 & 0 & 0 & 0 & 0 & 0 & 0 & 0  \\
            -4 L_9 & 0 & 0 & 0 & 0 & 0 & 0 & 0 & 0 & 0  \\
            2 L_{38} & 0 & 0 & 0 & 0 & 0 & 0 & 0 & 0 & 0  \\
            0 & 0 & 0 & 0 & 0 & 0 & 0 & 0 & 0 & 0  \\
            0 & -4 L_9 & \frac{3L_2}{2}-4 L_9 & 0 & 0 & 0 & 0 & 0 & 0 & 0  \\
            0 & 0 & \frac{L_2}{8} & 0 & 0 & 0 & 0 & 0 & 0 & 0  \\
            0 & 2 L_{38} & \frac{3L_{38}}{4} & -L_5 & 0 & 0 & 0 & 0 & 0 & 0  \\
            0 & 0 & -\frac{3L_{38}}{4} & 0 & 0 & 0 & 0 & 0 & 0 & 0  \\
            0 & 0 & 0 & \frac{L_{38}}{4} & 0 & 0 & 0 & 0 & 0 & 0  \\
            0 & 0 & 0 & 0 & -4 L_9 & \frac{3L_2}{2} -4 L_9 & -6 L_2 & 0 & 0 & 0  \\
            0 & 0 & 0 & 0 & 0 & \frac{L_2}{8} & -\frac{L_2}{2} & 0 & 0 & 0  \\
            0 & 0 & 0 & 0 & 2 L_{38} & \frac{3 L_{38}}{4} & 3 L_{38} & -L_5 & 0 & 0  \\
            0 & 0 & 0 & 0 & 0 & -\frac{3 L_{38}}{4} & -3 L_{38} & 0 & -L_5 & 3 L_{38}  \\
            0 & 0 & 0 & 0 & 0 & 0 & 0 & \frac{L_{38}}{4} & \frac{L_{38}}{4} & L_2  \\
            0 & 0 & 0 & 0 & 0 & 0 & -L_2 & 0 & -\frac{L_{38}}{8} & \frac{L_2}{2}  \\
            0 & 0 & 0 & 0 & 0 & 0 & 4 L_{38} & 0 & L_5 & -L_{38}  \\
            0 & 0 & 0 & 0 & 0 & 0 & 2 L_2 & 0 & 0 & 0  \\
        \end{pmatrix}
    \end{small}
,
\end{align}
\begin{align}
\mathbb{N}_{9\times17} = 
        \setcounter{MaxMatrixCols}{20}
\begin{small}
\begin{pmatrix}
\scriptstyle
             \frac{L_2}{8} & -\frac{L_2}{2} & 0 & 0 & 0 & 0 & 0 & 0 \\
            -\frac{3 L_{38}}{4}  & -3 L_{38} & 0 & -L_5 & 3 L_{38} & 0 & 0 & 0  \\
             0 & 0 & \frac{L_{38}}{4} & \frac{L_{38}}{4} & L_2 & 0 & 0 & 0  \\
             0 & -L_2 & 0 & -\frac{L_{38}}{8} & \frac{L_2}{2} & 0 & 0 & 0  \\
             0 & 4 L_{38} & 0 & L_5 & -L_{38} & -4 L_{38} & -2 L_5 & 2 L_{38}  \\
             0 & 2 L_2 & 0 & 0 & 0 & 0 & 0 & 0  \\
        \end{pmatrix}
\end{small}
,
\end{align}
and
\begin{align}
\mathbb{P}_{3\times9} = 
        \setcounter{MaxMatrixCols}{20}
\begin{small}
\begin{pmatrix}
            -L_2 & -\frac{L_{38}}{8} & \frac{L_2}{2} & 0 & 0 & 0 & 0 & 0 & 0 \\
            4 L_{38} & L_5 & -L_{38} & -4 L_{38} & -2 L_5 & 2 L_{38} & 0 & 0 & 0 \\
            2 L_2 & 0 & 0 & 0 & 0 & 0 & 0 & 0 & 0 \\
\end{pmatrix}
\end{small}
,
\end{align}
while $\mathbb{O}_{m \times n}$ are null $m \times n$ matrices.

To fix the integration constants, we match the solution of the differential equations to boundary values in the limit $s\to0$. All components of $\vec{w}$ are equal to zero except for
\begin{align}
    \left.w_{1_0}^{(0)}\right|_{s=0} = -1
    \,,\qquad\qquad\textup{and}\qquad\qquad
    \left.w_{1_0}^{(2)}\right|_{s=0} = -3\,\zeta_2
    \,.
\end{align}

 \section{Evaluation of Chen iterated integrals}
\label{app:cii}
 
To evaluate a Chen iterated integral $\left[\alpha_1,\dots\alpha_n\right]$ in the phase-space point $\vec{x}=\{\sbar{},\tbar{},\ubar{},1\}$, we consider a straight line from the large-mass limit $\vec{x}_0=\{0,0,0,1\}$ to $\vec{x}$ as integration path:
\begin{align}
    \gamma:(\vec{x}) \mapsto (1-\taumod{})\vec{x}_0 + \taumod{}\vec{x}\,.
    \label{eq:gamma}
\end{align}

We rewrite the integration kernels $L_i=\log\alpha_i$ in terms of the line parameter $\taumod{}\in[0,1]$, obtaining
    \begin{align}
    \left[\right]_{\vec{x}_0}\left(\vec{x}\right)
    &= 1
    \,,\notag\\
    \left[\alpha_{i_{1}}\right]_{\vec{x}_{0}}\left(\vec{x}\right)
    &= \int_0^1 \td\taumod{}_1\,f_1(\taumod{}_1)
    \,,\label{eq:deftau}\\
    \left[\alpha_{i_1},\hdots,\alpha_{i_{n-1}},\alpha_{i_n}\right]_{\vec{x}_0}(\vec{x})
    &= \int_0^1 \td\taumod{}_n\,f_n(\taumod{}_n)\int_0^{\taumod{}_n} \td\taumod{}_{n-1}\,f_{n-1}(\taumod{}_{n-1})\dots\int_0^{\taumod{}_2} \td\taumod{}_1\,f_1(\taumod{}_1)
    \,,\notag
    \end{align}
where the functions $f_i(\taumod{}_i)$ are the pullbacks of the differential forms $\dlog\,\alpha_i$ along the path $\gamma$: $\td\taumod{}_n\, f_n(\taumod{}_n) = \gamma^*\,\dlog\,\alpha_{i_n}$, i.e.\ $f_n(\taumod{}_n) = L_{i_n}(\vec{x}(\taumod{}_n))$.

The pullbacks $f_n(\taumod{}_n)$ might become degenerate along the path $\gamma$. In particular, they might:

\begin{itemize}
    \item \emph{Vanish}, in case no dependence over the line parameter is left. In this work, for general values of the kinematics, this happens only for the kernels $L_{31}$ and $L_{33}$. Due to the absence of the line parameter $\taumod{}$, all iterated integrals containing such kernels will evaluate to zero. Pullbacks might also vanish for specific phase-space points $\vec{x}$ (e.g., $L_{21}$ drops any dependence on $\taumod{}$ for phase-space points satisfying $\mbar_H^2 = \tbar^2/(\tbar-1)$).
    \item \emph{Collapse on a single term}, since overall constant factors will be lost through differentiation. In our calculation, this happens for
    \begin{align}
        \begin{aligned}
            L_1     &= \log m_H^2       &&\rightarrow \log \taumod{}    \,,\\
            L_2     &= \log s           &&\rightarrow \log \taumod{}    \,,\\
            L_3     &= \log t           &&\rightarrow \log \taumod{}    \,,\\
            L_4     &= \log u           &&\rightarrow \log \taumod{}    \,,\\
            L_7     &= \log (4m_H^2-s)  &&\rightarrow \log \taumod{}    \,,\\
            L_{12}  &= \log (m_H^2-t)   &&\rightarrow \log \taumod{}    \,,\\
            L_{13}  &= \log (m_H^2-u)   &&\rightarrow \log \taumod{}    \,,\\
            L_{19}  &= \log (m_H^4-t u) &&\rightarrow 2 \log \taumod{}  \,.
        \end{aligned}
    \end{align}
\end{itemize}

The integration kernels now might carry a \emph{starting point singularity}, which cannot be addressed using the general formula of Eq.~\eqref{eq:deftau} and require a dedicated prescription. This issue is analogous to the $G(0,\dots,0;x)$ term in the context of generalised polylogarithms (GPLs)~\cite{Goncharov:1998kja}, 
where it is defined as
\begin{align}
    G(0,\dots,0;\taumod{}) \equiv \frac{1}{n!}\log^n \taumod{}
    \,.
\end{align}
In our case, the problematic kernel is $\dlog\,\taumod{}$.
Hence, we define the iterated integral
\begin{align}
    \int_0^\taumod{} \dlog\,\taumod{}_n
    \int_0^{\taumod{}_{n}} \dlog\,\taumod{}_{n-1}
    \dots \int_0^{\taumod{}_2} \dlog\,\taumod{}_1 
    \equiv \frac{1}{n!}\log^n\taumod{}
    \,.
    \label{eq:logx}
\end{align}

When considering iterated integrals with general $\dlog$ integration kernels, some letters might contain an overall $\taumod{}^n$ factor in the argument of the logarithm. Such factor needs to be extracted to make integrals free of starting point singularities. In our case, we find:
\begin{align}
    \begin{aligned}
        L_{14} &\rightarrow   \log\taumod{} &&+ \log\left[\mbar{}_H^4 \sbar{} \taumod{}^2-3 \mbar{}_H^2 \sbar{} \taumod{}+\mbar{}_H^2+\sbar{}^2 \taumod{}\right]   ,\\
        L_{17} &\rightarrow   \log\taumod{} &&+ \log\left[\mbar{}_H^4 \taumod{}-4 \mbar{}_H^2+\sbar{}\right]   ,\\
        L_{20} &\rightarrow 2~\log\taumod{} &&+ \log\left[\sbar{} \tbar{} \ubar{} \taumod{}+\tbar{}^2-2 \tbar{} \ubar{}+\ubar{}^2\right]   ,\\
        L_{21} &\rightarrow   \log\taumod{} &&+ \log\left[-\mbar{}_H^2 \tbar{} \taumod{}+\mbar{}_H^2+\tbar{}^2 \taumod{}\right]   ,\\
        L_{22} &\rightarrow \log\taumod{} &&+   \log\left[-\mbar{}_H^2 \ubar{} \taumod{}+\mbar{}_H^2+\ubar{}^2 \taumod{}\right]   ,\\
        L_{23} &\rightarrow \log\taumod{} &&+   \log\left[\mbar{}_H^2-\sbar{} \tbar{} \taumod{}-\ubar{}\right]   ,\\
        L_{24} &\rightarrow \log\taumod{} &&+   \log\left[\mbar{}_H^2-\sbar{} \ubar{} \taumod{}-\tbar{}\right]   ,\\
        L_{25} &\rightarrow \log\taumod{} &&+   \log\left[\mbar{}_H^4 \taumod{}-\sbar{}-\tbar{} \ubar{} \taumod{}\right]   ,\\
        L_{27} &\rightarrow 2~\log\taumod{} &&+ \log\left[\mbar{}_H^4 \sbar{} \taumod{}+\mbar{}_H^4-\mbar{}_H^2 \sbar{} \ubar{} \taumod{}-2 \mbar{}_H^2 \tbar{}+\tbar{}^2\right]   ,\\
        L_{28} &\rightarrow 2~\log\taumod{} &&+ \log\left[\mbar{}_H^4 \sbar{} \taumod{}+\mbar{}_H^4-\mbar{}_H^2 \sbar{} \tbar{} \taumod{}-2 \mbar{}_H^2 \ubar{}+\ubar{}^2\right]   ,
    \end{aligned}
\end{align}
where the logarithms on the r.h.s.\ are now regular at $\taumod{}\to0$, and we took $m_V^2=1$ for simplicity.

Another delicate point is the presence of \emph{endpoint singularities}, that is, singularities occurring in the outermost integration kernel at $\taumod{}=1$, resulting in a divergent iterated integral. Except for $L_{31}$, $L_{33}$ (which do not contain $\taumod{}$ at all), $L_{35}$, $L_{36}$, $L_{37}$, $L_{38}$, $L_{41}$, $L_{43}$, $L_{48}$, and $L_{49}$, all integration kernels can develop endpoint singularities. Such singularities can be extracted applying shuffle relations to the iterated integrals, even though a much more practical solution is to add a small offset to the phase-space point of interest.

Once we have made sure that the iterated integrals are all finite, except for some explicit singular terms, we can proceed to numerically evaluate them. To this end, we employ the fact that we can differentiate a depth $n$ iterated integral $n$ times w.r.t.\ the line parameter $\taumod{}$, obtaining a closed system of differential equations of the form
\begin{align}
    \begin{aligned}
        \partial_\taumod{} \left[\alpha_{i_1},\dots,\alpha_{i_{n-1}},\alpha_{i_n}\right]_0(\taumod{}) &= \left(\partial_\taumod{} \alpha_{i_n}(\taumod{})\right) \left[\alpha_{i_1},\dots,\alpha_{i_{n-1}}\right]_0(\taumod{})  \,,\\
        &~~\vdots  \\
        \partial_\taumod{} \left[\alpha_1\right]_0(\taumod{}) &= \left(\partial_\taumod{} \alpha_1(\taumod{})\right) \left[\right]_0(\taumod{}) \,,\\
        \partial_\taumod{} \left[\right]_0(\taumod{}) &= 0    \,,
    \end{aligned}
\end{align}
where the notation is to be understood as
\begin{align}
    \left[\alpha_{i_1},\dots,\alpha_{i_{n-1}},\alpha_{i_n}\right]_0(\taumod{})
    \equiv
    \int_0^\taumod{} \td\taumod{}_n\,f_{i_n}(\taumod{}_n)\int_0^{\taumod{}_n} \td\taumod{}_{n-1}\,f_{i_{n-1}}(\taumod{}_{n-1})\dots\int_0^{\taumod{}_2} \td\taumod{}_1\,f_{i_1}(\taumod{}_1)\,,
\end{align}
with boundary conditions in $\taumod{}=0$
\begin{align}
    \begin{aligned}
        \left[\alpha_1,\dots,\alpha_{n-1},\alpha_n\right]_0(0) &= 0  \,,\\
        &~~\vdots  \\
        \left[\alpha_1\right]_0(0) &= 0 \,,\\
        \left[\right]_0(0) &= 1 \,.
    \end{aligned}
\end{align}

We rely on the \textsc{Mathematica} package \textsc{DiffExp} \cite{Hidding:2020ytt} to implement the discussion above and numerically evaluate the iterated integrals. To make full use of \textsc{DiffExp} routines we also need to provide all imaginary prescriptions ($+\iu\delta$ or $-\iu\delta$) necessary to cross the singularities along the integration path. The individual imaginary prescriptions are uniquely determined from the Feynman prescription on the Mandelstam invariants ($+\iu\delta$), except for $\log\taumod{}$, which we set to $+\iu\delta$. This arbitrary choice will not introduce any ambiguity in the numerical results of the functions $\mathcal{A}_i^{(2)}$.

 \section{Computational details}
\label{app:computation}
The computation time required to evaluate the independent functions is strongly influenced by the size of the numerators and denominators at a given point, as {\sc DiffExp} performs significantly faster when handling ``simple" rational numbers.
To illustrate this feature, let us consider the following two points: 
\begin{align}
x_1  & =\left\{ \frac{3125}{128}\,,-\frac{1875}{128}\,,-\frac{625}{128}\,,\frac{625}{256}\,,1\right\}\,, \\ 
x_2  & =\left\{ \frac{3125}{128}+10^{-15}\,,-\frac{1875}{128}+10^{-15}\,,-\frac{625}{128}+10^{-15}\,,\frac{625}{256}\,,1\right\}\,,
\end{align}
with $x_i = \{s_i,\, t_i,\, u_i,\, m_{H,i}^2,\, m_{V, i}^2\}$. 

We evaluate the functions at both points by solving the differential equations in a straight-line path starting from the large mass expansion.
We target two orders in $\epsilon$ and a numerical precision of $16$ significant figures. Transporting the solution to point $x_1$ requires $27$ integration segments and a total computation time of $218$ seconds.
In contrast, evaluation at point $x_2$ requires $337$ segments in $1809$ seconds.

Let us also remark that the requested precision is not always guaranteed by {\sc DiffExp}. A pratical method of estimating the actual numerical uncertainty is to evaluate a given point along two distinct integration paths. As an alternative path, we consider the straight line starting from the point $x_0$ defined in Eq.~\eqref{eq:psp_test}. The evaluation takes $39$ seconds for $x_1$ and $120$ seconds for $x_2$. 
By comparing the numerical evaluation of the independent functions (as well as the form factors),
obtained from both paths, we estimate their numerical uncertainty to be of the orders $\mathcal{O}\left(10^{-29}\right)$ for $x_1$ and $\mathcal{O}\left(10^{-12}\right)$ for $x_2$. 

The benchmarks were obtained with a single thread on a system equipped with an {\sc AMD Ryzen 9 9700X} and $32$ GB of DDR5 RAM, running {\sc Ubuntu 24.04}.

 \section{Organisation of ancillary files}
\label{app:files}

The ancillary files of this paper can be found at~\cite{zenodo}. 
They are split into two main directories containing the analytic construction of the canonical integrals and of the form factors. 
We describe below the files contained in each directory.

\subsection*{Canonical integrals}

\begin{itemize}
\item \verb"Alphabet.m": includes the letters of the alphabet of the integral families discussed in this paper. Each letter is represented by \verb"a[i]" with \verb"i"$\,=0,1,\dots,77$.

\item \verb"Sqrts.m": contains the definition of the $r_i$ of Eq.~\eqref{eq:sqrts} 
in terms of irreducible square roots, suitable for implementation with {\sc DiffExp} (e.g.\ $r_1=\sqrt{s}\sqrt{s-4m_{H}^{2}}$). 

\item \verb"Atilde_fam.m" (with \verb"fam"$\,=\,$\verb"PL", \verb"PLx12", \verb"NP"
): contains the connection matrices $\tilde{A}_{\TOP[X]}$ with  $\TOP[X] = \PL\,, \PLx\,,  \NP$,
according to Eq.~\eqref{eq:deq_can}.
\item \verb"MIs_fam.m" (with \verb"fam"$\,=\,$\verb"PL", \verb"PLx12", \verb"NP"
): contains the canonical master integrals 
$\vec{J}_{\TOP[X]}$ with  $\TOP[X] = \PL\,, \PLx\,,  \NP$.
The definition of the integral follows the notation of Eq.~\eqref{eq:Feyn_int}.
\item\verb"Mappings_J_to_J.m": contains the sector mappings among the canonical integrals of families $\PL$, $\PLx$, and $\NP$.
\item\verb"Mappings_J_to_W.m": contains the mappings of the canonical integrals of $\PL$, $\PLx$,  and $\NP$ onto the rotated canonical integrals $\vec{W}$ of Eq.~\eqref{eq:rot_wfuns}.
ash\item\verb"Mappings_W_to_w.m": contains the decomposition of the rotated canonical integrals $\vec{W}$ into transcendental functions $w_{i_k}^{(k')}$, 
according to Eq.~\eqref{eq:ind_wfun}.
\item\verb"Solution_w_CII.m": contains the solution of the functions $w_{i_k}^{(k')}$ up to transcendental weight six in terms of Chen iterated integrals. 
\item\verb"DEQ_w.m": contains the differential equations (\verb"DEQ") for all the 327  functions $w_{i_k}^{(k')}$ (\verb"wfun") 
functions present in the analytic evaluation of the rotated canonical integrals $\vec{W}$, 
and the boundary values at $s=t=u=0$, coming from $s>0$ (\verb"wfun0").

\end{itemize}

\subsection*{Form factors}

\begin{itemize}

\item\verb"Ai.m" (with $i=1,2,3$): contains the functions $\mathcal{A}_{i;n}^{(2)}$ (\verb"Ai[n]") at order $\mathcal{O}\left(\epsilon^n\right)$ with $n=0,1,2$.

\item\verb"run_DiffExp.m": contains the script to numerically evaluate the functions \\$w_{i_k}^{(k')}$ and $\mathcal{A}_{i;n}^{(2)}$. 

\end{itemize}

\bibliographystyle{JHEP}
\bibliography{Biblio}

\end{document}